\documentclass{article}
\usepackage{amsmath}
\usepackage{amssymb}
\usepackage{graphicx}
\usepackage{color}

\begin{document}

\title{Stabilization of three-wave vortex beams in the waveguide}
\author{Arnaldo Gammal$^{1}$ and Boris A. Malomed$^{2}$ \\
$^{1}$Instituto de F\'{\i}sica, Universidade de S\~{a}o Paulo, 05508-090,
S\~{a}o Paulo, Brazil\\
$^{2}$Department of Physical Electronics, School of Electrical Engineering,\\
Faculty of Engineering, Tel Aviv University, Tel Aviv 69978, Israel}
\maketitle
\begin{center}{\bf Abstract}
\end{center}
We consider two-dimensional (2D) localized vortical modes in the three-wave
system with the quadratic ($\chi ^{(2)}$) nonlinearity, alias nondegenerate
second-harmonic-generating system, guided by the isotropic
harmonic-oscillator (HO)
(alias parabolic)
confining potential. In addition to the
straightforward realization in optics, the system models mixed
atomic-molecular Bose-Einstein condensates (BECs). The main issue is
stability of the vortex modes, which is investigated through computation of
instability growth rates for eigenmodes of small perturbations, and by means
of direct simulations. The threshold of parametric instability for
single-color beams, represented solely by the second harmonic (SH) with zero
vorticity, is found in an analytical form with the help of the variational
approximation (VA). Trapped states with vorticities $\left( +1,-1,0\right) $
in the two fundamental-frequency (FF) components and the SH one [the
so-called \textit{hidden-vorticity} (HV) modes] are completely unstable.
Also unstable are \textit{semi-vortices} (SVs), with component vorticities $%
\left( 1,0,1\right) $. However, full vortices, with charges $\left(
1,1,2\right) $, have a well-defined stability region. 
Unstable full vortices feature regions of robust dynamical behavior, where
they periodically split and recombine, keeping their vortical content.
\bigskip

 \bigskip

\section{Introduction}

The fundamental significance of the quadratic ($\chi ^{(2)}$) nonlinearity
in optics, including its use for the creation of solitons, is well known
\cite{review0}-\cite{review}, \cite{KA}. In particular, the $\chi ^{(2)}$
nonlinearity opens the way to the making of two- and three-dimensional (2D
and 3D) solitons, because, unlike the Kerr (cubic) terms, the quadratic
ones, which couple the fundamental-frequency (FF) and second-harmonic (SH)
components of the optical fields, do not give rise to the collapse in the
two- and three-dimensional (2D and 3D) space\ \cite{Rubenchik}, which is a
severe problem for multidimensional solitons in Kerr media \cite{review}.
Stable (2+1)D beams propagating in quadratically nonlinear optical media
have been demonstrated in the experiment \cite{first-exper}, and stable
spatiotemporal ``light bullets" have been predicted in these
media too \cite{Drummond}. In the experiment, fully self-trapped
``bullets" have not been reported yet, the closest result
being a spatiotemporal soliton self-trapped in the longitudinal and one
transverse directions, while the confinement in the additional transverse
direction was imposed by the guiding structure \cite{Wise1,Wise2}.

A challenging issue for the (2+1)D setting is the search for conditions
securing the stability of vortical solitons. In contrast to their
fundamental counterparts, vortex solitons supported by the quadratic
instability in the free space are always unstable against azimuthal
perturbations, which tend to split the vortex into a set of separating
fragments. For the degenerate quadratic system, with a single FF component,
the splitting instability was predicted theoretically \cite{splitting1}-\cite%
{splitting5} and demonstrated in the experiment \cite{splitting-exp}. The
same instability occurs in the framework of the full three-wave system,
which includes two distinct FF components, which represent mutually
orthogonal polarizations of light \cite{splitting-3W,splitting-3W2}.

Vortex solitons are also known as solutions of the 2D nonlinear Schr\"{o}%
dinger (NLS) equation with the self-focusing cubic term \cite{Kruglov}. They
too are \ subject to the azimuthal instability, which is actually stronger
than the collapse instability driven by the cubic nonlinearity \cite{review}%
. The 2D NLS equation models not only light beams in bulk media with the
Kerr nonlinearity, but also the mean-field dynamics of atomic Bose-Einstein
condensates (BECs), trapped in the form of ``pancakes" by
the confining potential \cite{BEC}; in the latter case, the NLS equation is
usually called the Gross-Pitaevskii (GP) equation. In the context of the
NLS/GP equation, a practically relevant solution of the instability problem
was elaborated: both fundamental and vortical solitons, with topological
charge $m=0$ and $1$, respectively, can be stabilized by 2D
harmonic-oscillator (HO) confining potentials. It is well known \cite%
{cubic-in-trap1}-\cite{cubic-in-trap7} that the HO potential renders the
fundamental solitons stable, against small perturbations, in the entire
region of their existence. The vortices with $m=1$, trapped by the same
potential, are stabilized in $\simeq 33\%$ of their existence region (in
terms of their norm). Additionally, an adjacent region of width $\simeq 10\%$
supports a robust dynamical regime featuring periodic splitting and
recombination of the vortices, which keep their topological charge \cite%
{cubic-in-trap5}. As concerns the BEC realization, stabilization of
self-trapped semi-vortex modes in the free 2D space was recently
demonstrated in a two-component system with the Kerr nonlinearity and linear
spin-orbit coupling between the components \cite{semi}.

The effective 2D trapping potential can be also realized in optical
waveguides, in the form of the respective profile of the transverse
modulation of the local refractive index \cite{KA}. This circumstance
suggests a natural possibility for the stabilization of (2+1)D vortex
solitons in the $\chi ^{(2)}$ medium by means of the radial HO potential.
For the degenerate (two-wave) $\chi ^{(2)}$ system, alias the \textit{Type-I}
second-harmonic-generating interaction \cite{review0}-\cite{review2}, this
possibility was demonstrated in Ref. \cite{HS}, and then extended for 3D
spatiotemporal vortex solitons in Ref. \cite{HS-3D}. The subject of the
present work is to investigate this stabilization mechanism for the
nondegenerate three-wave system, with the \textit{Type-II} quadratic
interaction, where the results demonstrate essentially new features in
comparison with the degenerate model.

A feasible approach to the making of the optical medium combining a
nearly-parabolic profile of the refractive index and $\chi ^{(2)}$
nonlinearity is the use of a 2D photonic crystal, which can be readily
designed to emulate the required index profile, while the quadratic
nonlinearity may be provided by poled material filling voids of the
photonic-crystal matrix \cite{Du}, \cite{Luan}. In fact, the effective
radial potential provides for sufficiently strong localization of the
trapped modes, therefore the exact parabolic shape of the radial profile is
not crucially important \cite{HS}.

Essentially the same system of GP equations for atomic and molecular wave
functions models the BEC in atomic-molecular mixtures \cite{BEC0}-\cite{BEC4}%
. In that case, two different components of the atomic wave function pertain
to two different atomic states of the same species, while the quadratic
nonlinearity accounts for reactions of the merger of two atoms into a
molecule, and splitting of the molecules due to collisions with free atoms.
Accordingly, the predicted mechanism of the stabilization of three-component
vortex solitons trapped in the HO potential can also be realized in the BEC
mixture.

The paper is organized as follows. The model, based on the system of three
propagation equation coupled by the $\chi ^{(2)}$ terms, is introduced in
Section II. In Section III, we introduce four types of three-wave modes
considered in this work: the SH-only (single-color) one, with zero FF
components; modes with the \textit{hidden vorticity} (HV, which means
opposite vorticities, $S=\pm 1$, in the two FF components, and $S=0$ in the
SH\ component); \textit{semi-vortices}, with $S=0$ in one FF component and $%
S=1$ in the other one and in the SH wave (cf. the above-mentioned
semi-vortices in the two-component system with the cubic nonlinearity and
linear spin-orbit coupling between the components \cite{semi}); and,
finally, \textit{full vortices}, with $S=1$ in both FF components, and $S=2$
in the SH. In Section IV we report simple but nontrivial analytical results,
which predict the threshold of the parametric instability for the
fundamental and vortical single-color states by means of the variational
approximation (VA). In Section V we formulate the eigenvalue problem for
small perturbations around stationary solutions, which determines their
stability. The numerical results for modes of all the types are reported in
Section VI. First, we identify the stability threshold for the fundamental
and vortical single-color trapped modes, and compare these results with the
above-mentioned analytical predictions provided by the VA. Next, the results
are summarized for the HV and SV modes, which are found to be unstable.
Finally, the stability region is presented for the most general three-wave
states, with vorticities $1$ in both FF components and $2$ in the SH. In
addition, unstable vortices may exist in a quasi-stable dynamical regime, in
the form of periodic splitting and recombination of the semi-vortex. The
paper is concluded by Section VII.

\section{The model}

The copropagation of two components of the FF wave, with amplitudes $u\left(
x,y,z\right) $ and $v\left( x,y,z\right) $, and the SH wave, with amplitude $%
w\left( x,z\right) $, in the bulk $\chi ^{(2)}$ waveguide obeys the system
of three coupled equations written here in the normalized form \cite%
{review1,review2}:
\begin{eqnarray}
iu_{z} &=&-\frac{1}{2}\nabla ^{2}u+v^{\ast }w+\frac{1}{2}\Omega
^{2}r^{2}u+Qu,  \label{u} \\
iv_{z} &=&-\frac{1}{2}\nabla ^{2}v+u^{\ast }w+\frac{1}{2}\Omega
^{2}r^{2}v-Qv,  \label{v} \\
2iw_{z} &=&-\frac{1}{2}\nabla ^{2}w-qw+uv+2\Omega ^{2}r^{2}w,  \label{w}
\end{eqnarray}%
where $\nabla ^{2}\equiv \partial ^{2}/\partial x^{2}+\partial ^{2}/\partial
y^{2}$ is the transverse diffraction operator, the asterisk stands for the
complex conjugate, $Q$ is the birefringence coefficient, $q$ is the FF-SH
phase mismatch, and $\Omega ^{2}$ the strength of the HO trapping potential.
In the most general case, stationary vortex solutions of this system, with
two independent FF propagation constants $k_{u}$ and $k_{v}$, can be looked
for as
\begin{eqnarray}
\left\{ u,v\right\}  &=&\left\{ U_{0}(r)e^{iS_{u}\theta
+ik_{u}z},V_{0}(r)e^{iS_{v}\theta +ik_{v}z}\right\} ,  \label{UV} \\
w &=&W_{0}(r)e^{i\left( S_{u}+S_{v}\right) +i\left( k_{u}+k_{v}\right) z},
\label{SH}
\end{eqnarray}%
where $S_{u}$ and $S_{v}$ are integer vorticities of the two FF components,
and $U_{0}(r)$, $V_{0}(r)$, and $W_{0}(r)$ are real radial functions with
asymptotic forms%
\begin{gather}
\left\{ U_{0}(r),V_{0}(r),W_{0}(r)\right\} \sim r^{\left\{ \left\vert
S_{u}\right\vert ,\left\vert S_{v}\right\vert ,\left\vert
S_{u}+S_{v}\right\vert \right\} }~\ \mathrm{at}~~r\rightarrow 0,  \notag \\
\left\{ U_{0}(r),V_{0}(r)\right\} \sim \exp \left( -\frac{\Omega }{2}%
r^{2}\right) ,~W_{0}(r)\sim \exp \left( -\Omega r^{2}\right) \ \mathrm{at}%
~~r\rightarrow \infty .  \label{asympt}
\end{gather}
The exact equality of the propagation constant and vorticity of the SH component
in Eqs. (\ref{UV}) and (\ref{SH}) to sums of the same constants of the
FF components is imposed by the coherent coupling between the FF and SH components
in Eqs. (\ref{u}), (\ref{v}), and (\ref{w}).

In this work, the analysis is carried out, chiefly, for $Q=0$, in which case
the symmetry between the two FF components suggests to consider three-wave
states with $k_{u}=k_{v}$. For $Q\neq 0$, rescaling of Eqs. (\ref{u})-(\ref%
{v}) makes it possible to fix $Q=1$. Families of three-wave vortices with $%
Q=1$ are considered too, at the end of the paper.

Equations (\ref{u})-(\ref{w}) can be derived from the Lagrangian,%
\begin{equation}
L=\int \int \left\{ \frac{i}{2}\left[ \left( uu_{z}^{\ast }-u^{\ast
}u_{z}\right) +i\left( vv_{z}^{\ast }-v^{\ast }v_{z}\right) +2\left(
ww_{z}^{\ast }-w^{\ast }w_{z}\right) \right] \right\} dxdy-H,  \label{L}
\end{equation}%
where the Hamiltonian is%
\begin{eqnarray}
H &=&\int \int \left[ \frac{1}{2}\left( \left\vert \nabla u\right\vert
^{2}+\left\vert \nabla u\right\vert ^{2}+\left\vert \nabla w\right\vert
^{2}\right) +\frac{1}{2}\Omega ^{2}r^{2}\left(
|u|^{2}+|v|^{2}+4|w|^{2}\right) \right.  \notag \\
&&\left. +Q\left( |u|^{2}-|v|^{2}\right) -q|w|^{2}+\left( uvw^{\ast
}+u^{\ast }v^{\ast }w\right) \right] dxdy.  \label{H}
\end{eqnarray}

The Hamiltonian is the dynamical invariant of Eqs. (\ref{u})-(\ref{w}),
along with the total power (field norm),%
\begin{equation}
N=\int \int \left( |u|^{2}+|v|^{2}+4|w|^{2}\right) dxdy\equiv
N_{u}+N_{v}+N_{w},  \label{N}
\end{equation}%
and the total angular momentum,%
\begin{equation}
M=i\int \int \left[ u^{\ast }\left( yu_{x}-xu_{y}\right) +v^{\ast }\left(
yv_{x}-xv_{y}\right) +2w^{\ast }\left( yw_{x}-xw_{y}\right) \right] dxdy,
\label{M}
\end{equation}%
which is a real integral quantity, even if it formally seems complex.
(it can be checked by means of the integration by parts that the $M^{*}\equiv M$).

\section{Stationary solutions}

\subsection{Single-color (second-harmonic) states}

The simplest stationary state in the present model is the one with unexcited
FF components, i.e., $U_{0}=V_{0}=0$ in Eq. (\ref{UV}) and $W_{0}(r)$ from
Eq. (\ref{SH}) obeying the linear equation, which follows from Eq. (\ref{w}%
):
\begin{equation}
\left( -4k+q\right) W_{0}=-\frac{1}{2}\left( W_{0}^{\prime \prime }+\frac{1}{%
r}W_{0}^{\prime }-\frac{S_{w}^{2}}{r^{2}}\right) +2\Omega ^{2}r^{2}W_{0},
\label{linear}
\end{equation}%
where $S_{w}$ replaces $\left( S_{u}+S_{v}\right) $, see Eq. (\ref{SH}).
This equation is tantamount to the radial Schr\"{o}dinger equation for 2D
quantum-mechanical states with azimuthal quantum number $S$ in the HO
potential. The respective solutions to Eq. (\ref{linear}) are%
\begin{gather}
W_{0}=Br^{|S_{w}|}\exp \left( -\Omega r^{2}\right) ,  \label{W0} \\
k=-\Omega \left( 1+|S_{w}|\right) /2+q/4,  \label{k}
\end{gather}%
where $B$ is an arbitrary constant. The norm (\ref{N}) of this state is%
\begin{equation}
N_{\mathrm{SH}}=\frac{4\pi \left\vert S_{w}\right\vert !}{\left( 2\Omega
\right) ^{1+\left\vert S_{w}\right\vert }}B^{2}.  \label{NSH}
\end{equation}

\subsection{Hidden-vorticity (HV) states}

Solutions for 2D HV modes are defined by the vorticity set $\left(
S_{u},S_{v},S_{w}\equiv S_{u}+S_{v}\right) =\left( +1,-1,0\right) $ of the
three components in Eqs. (\ref{UV}) and (\ref{SH}). Accordingly, the
solutions are looked for as%
\begin{gather}
\left\{ u,v\right\} =\left\{ U_{0}(r)e^{iS\theta },V_{0}(r)e^{-iS\theta
}\right\} e^{ikz},  \label{uv} \\
w=W_{0}(r)e^{2ikz},  \label{w2}
\end{gather}%
where real functions $U_{0},$ $V_{0},$ and $W_{0}$ satisfy the following
radial equations (where the birefringence terms are kept, for the time
being):%
\begin{eqnarray}
-kU_{0} &=&-\frac{1}{2}\left( U_{0}^{\prime \prime }+\frac{1}{r}%
U_{0}^{\prime }-\frac{S^{2}}{r^{2}}U_{0}\right) +V_{0}W_{0}+\frac{1}{2}%
\Omega ^{2}r^{2}U_{0}+QU_{0},  \label{U} \\
-kV_{0} &=&-\frac{1}{2}\left( V_{0}^{\prime \prime }+\frac{1}{r}%
V_{0}^{\prime }-\frac{S^{2}}{r^{2}}V_{0}\right) +U_{0}W_{0}+\frac{1}{2}%
\Omega ^{2}r^{2}V_{0}-QV_{0},  \label{V} \\
-4kW_{0} &=&-\frac{1}{2}\left( W_{0}^{\prime \prime }+\frac{1}{r}%
W_{0}^{\prime }\right) -qW_{0}+U_{0}V_{0}+2\Omega ^{2}r^{2}W_{0}.  \label{W}
\end{eqnarray}

Previously, the HV concept was realized for the three-wave $\chi ^{(2)}$
system in the free space \cite{Lluis}, as well as for various systems of
coupled continuous \cite{hidden0}-\cite{hidden} and discrete \cite%
{Panos,Leykam} NLS/GP equations with the cubic nonlinearity, including the
system supported by the HO trapping potential \cite{hidden}.
In work \cite{Fangwei},
it was demonstrated that the HV essentially suppresses the modulational instability of
2D ring solitons in a two-component system with saturable nonlinearity, in comparison
with their counterparts carrying explicit vorticity. However, \textit{trapped} vortical
modes with HV were not previously studied in three-wave systems.

\subsection{Semi-vortices}

The other type of the vortical mode, with zero vorticity in one FF component
(hence they are called semi-vortices, as said above), is defined by the set
of $\left( S_{u},S_{v},S_{w}\right) =\left( 1,0,1\right) $, i.e., solution
ansatz (\ref{UV}), (\ref{SH}) reduces to%
\begin{eqnarray}
\left\{ u,v\right\} &=&\left\{ U_{0}(r)e^{i\theta
+ik_{1}z},V_{0}(r)e^{ik_{2}z}\right\} ,  \label{10} \\
w &=&W_{0}(r)e^{i\theta +i\left( k_{1}+k_{2}\right) z},  \label{1}
\end{eqnarray}%
where the real radial amplitudes obey the following equations, cf. Eqs. (\ref%
{U}), (\ref{V}), (\ref{W}):
\begin{eqnarray}
-k_{1}U_{0} &=&-\frac{1}{2}\left( U_{0}^{\prime \prime }+\frac{1}{r}%
U_{0}^{\prime }-\frac{1}{r^{2}}U_{0}\right) +V_{0}W_{0}+\frac{1}{2}\Omega
^{2}r^{2}U_{0}+QU_{0},  \notag \\
-k_{2}V_{0} &=&-\frac{1}{2}\left( V_{0}^{\prime \prime }+\frac{1}{r}%
V_{0}^{\prime }\right) +U_{0}W_{0}+\frac{1}{2}\Omega ^{2}r^{2}V_{0}-QV_{0},
\label{1-0-1} \\
-2\left( k_{1}+k_{2}\right) W_{0} &=&-\frac{1}{2}\left( W_{0}^{\prime \prime
}+\frac{1}{r}W_{0}^{\prime }-\frac{1}{r^{2}}W_{0}\right)
-qW_{0}+U_{0}V_{0}+2\Omega ^{2}r^{2}W_{0}.  \notag
\end{eqnarray}%
In the limit of small amplitudes, solutions of the linearized version of
Eqs. (\ref{1-0-1}) can be found in the form of
\begin{eqnarray}
U_{0} &=&U_{0}^{(0)}r\exp \left( -\Omega r^{2}/2\right) ,  \notag \\
V_{0} &=&W_{0}^{(0)}\exp \left( -\Omega r^{2}/2\right) ,  \label{000} \\
W_{0} &=&W_{0}^{(0)}r\exp \left( -\Omega r^{2}\right) ,  \notag
\end{eqnarray}%
where the propagation constant must satisfy, respectively, the following
relations for the components $U,V,W$:
\begin{eqnarray}
k_{1} &=&-2\Omega -Q,  \notag \\
k_{2} &=&-\Omega +Q,  \label{k-k-k} \\
k_{1}+k_{2} &=&-2\Omega +q/2,  \notag
\end{eqnarray}%
From Eqs. (\ref{k-k-k}) it follows that, in the small-amplitude limit, the
semi-vortex mode exists at the single value of the mismatch, $q=-2\Omega $.
Below, it is shown that solutions for nonlinear semi-vortex modes lift this
constraint.

\subsection{Full vortices}

In fact, the HV modes and semi-vortices are shown below to be completely
unstable (modes of the latter type may be replaced by robust dynamical
states which keep the same vortical structure). A stationary topological
state which features a well-defined stability region is the \textit{full
vortex}, with the set of $\left( S_{u},S_{v},S_{w}\right) =\left(
1,1,2\right) $ in Eqs. (\ref{UV}) and (\ref{SH}). The respective system of
stationary equations is [cf. Eqs. (\ref{U})-(\ref{W}) and (\ref{1-0-1})]:
\begin{eqnarray}
-k_{1}U_{0} &=&-\frac{1}{2}\left( U_{0}^{\prime \prime }+\frac{1}{r}%
U_{0}^{\prime }-\frac{1}{r^{2}}U_{0}\right) +V_{0}W_{0}+\frac{1}{2}\Omega
^{2}r^{2}U_{0}+QU_{0},  \notag \\
-k_{2}V_{0} &=&-\frac{1}{2}\left( V_{0}^{\prime \prime }+\frac{1}{r}%
V_{0}^{\prime }-\frac{1}{r^{2}}V_{0}\right) +U_{0}W_{0}+\frac{1}{2}\Omega
^{2}r^{2}V_{0}-QV_{0},  \label{1-1-2} \\
-2\left( k_{1}+k_{2}\right) W_{0} &=&-\frac{1}{2}\left( W_{0}^{\prime \prime
}+\frac{1}{r}W_{0}^{\prime }-\frac{4}{r^{2}}W_{0}\right)
-qW_{0}+U_{0}V_{0}+2\Omega ^{2}r^{2}W_{0}.  \notag
\end{eqnarray}

\section{The variational approximation (VA) for the onset of the parametric
instability of the single-color modes}

The VA is a natural approach for predicting the shape of nonlinear modes,
including vortical ones supported by the $\chi ^{(2)}$ nonlinearity \cite%
{HS,semi}. In the present context, the VA can be used for the prediction of
the instability threshold of the trapped single-color (SH-only) states (the
VA may be developed for other situations too, but it then turns out to be
rather cumbersome).

For this purpose, it is necessary to consider a perturbed version of the
single-color stationary state, which includes the FF components too. In
particular, in the case of zero vorticity of the SH component and
vorticities $\pm S$ of the FF perturbations, the Lagrangian of the
respective stationary equations (\ref{U})-(\ref{W}) can be reduced to its
radial part:%
\begin{gather}
L_{\mathrm{rad}}=\int_{0}^{\infty }rdr\left\{ \frac{1}{2}\left[ \left( \frac{%
dU_{0}}{dr}\right) ^{2}+\left( \frac{dV_{0}}{dr}\right) ^{2}+\left( \frac{%
dW_{0}}{dr}\right) ^{2}+\frac{S^{2}}{r^{2}}\left( U_{0}^{2}+V_{0}^{2}\right) %
\right] \right.  \notag \\
+k\left( U_{0}^{2}+V_{0}^{2}+4W_{0}^{2}\right) +\frac{1}{2}\Omega
^{2}r^{2}\left( U_{0}^{2}+V_{0}^{2}+4W_{0}^{2}\right) +Q\left(
U_{0}^{2}-V_{0}^{2}\right) -qW_{0}^{2}  \notag \\
\left. +2U_{0}V_{0}W_{0}\right\} .  \label{Lrad}
\end{gather}%
To develop the VA, the simplest Gaussian ansatz may be adopted, assuming
equal amplitudes ($A$) and widths of fields $U_{0}$ and $V_{0}$:%
\begin{equation}
U_{0}(r)=V_{0}(r)=Ar^{S}\exp \left( -\Omega r^{2}/2\right) ,W_{0}(r)=B\exp
\left( -\Omega r^{2}\right)  \label{ans}
\end{equation}%
($S\geq 0$ is defined here). The symmetry between the FF\ components implies
that we should set $Q=0$, otherwise the symmetry will be broken by the
birefringence. Setting $Q=0$ limits the consideration to a particular case
of the generic three-wave system, but even this particular case is a
nontrivial one, as it has no counterpart in the usually considered
degenerate two-wave system, in which the VA was applied before to the
description of trapped vortices \cite{HS}. The widths of all the components
in ansatz (\ref{ans}) are not treated as variational parameters, but are
rather taken as per wave functions of states trapped in the 2D
harmonic-oscillator potential, cf. Eq. (\ref{W0}).

The objective of the use of the VA in the present context is to predict a
point at which a solution with an infinitely small $A$ appears, against the
background of the fundamental single-color (SH-only) state, given by Eqs. (%
\ref{W0}), (\ref{k}) with $S_{w}=0$ and arbitrary amplitude $B$. The
appearance of this mode signals the onset of the parametric instability of
the latter state \cite{HS}. The substitution of ansatz (\ref{ans}) into
radial Lagrangian (\ref{Lrad}) yields the following reduced Lagrangian, in
which we drop terms that produce no contribution in the subsequent analysis:%
\begin{equation}
L_{\mathrm{rad}}=S!\frac{A^{2}}{\Omega ^{S}}\left[ 1+S+\frac{k}{\Omega }%
+2^{-\left( 1+S\right) }\frac{B}{\Omega }\right] .  \label{Leff}
\end{equation}%
This Lagrangian gives rise to the Euler-Lagrange equation, $\partial L_{%
\mathrm{rad}}/\partial \left( A^{2}\right) =0$. The parametric-instability
threshold, corresponding to the emergence of the solution with an
infinitesimal $A^{2}>0$, corresponds to setting $A^{2}=0$ in the resulting
equation:%
\begin{equation}
B=-2^{1+S}\left[ \left( 1+S\right) \Omega +k\right] =-2^{1+S}\left[ \left(
\frac{1}{2}+S\right) \Omega +\frac{q}{4}\right] ,  \label{B}
\end{equation}%
where expression (\ref{k}) was substituted for $k$. The respective critical
total power (norm) of the trapped single-color state is given by Eq. (\ref%
{NSH}),%
\begin{equation}
N_{c}=2^{1+2S}\frac{\pi }{\Omega }\left[ \left( 1+2S\right) \Omega +\frac{q}{%
2}\right] ^{2}.  \label{Nc}
\end{equation}%
Thus, the fundamental single-color state is stable at $N<N_{c}$, and
unstable at $N>N_{c}$.

Note that, as it follows from Eq. (\ref{Nc}), the parametric instability of
the zero-vorticity single-color state is dominated by the vortical FF
perturbations with $S=1$ (i.e., $S=1$ give rise to $N_{c}$ \emph{lower }than
its counterpart produced by the zero-vorticity perturbations, with $S=0$) in
an interval of negative values of the mismatch, $\left( 14/3\right) \Omega
<-q<10\Omega $.

It is also relevant to develop a similar analysis for the onset of the
parametric instability of the vortical single-color trapped state with even
vorticity $S_{w}>0$, for which $W_{0}(r)=Br^{S_{w}}\exp \left( -\Omega
r^{2}\right) $, cf. Eq. (\ref{W0}), and $S$ is replaced by $S_{w}/2$ in the
FF components of ansatz (\ref{ans}). In this case, the reduced radial
Lagrangian (\ref{Leff}) is replaced by%
\begin{equation}
L_{\mathrm{rad}}^{\mathrm{(}S_{w}\mathrm{)}}=\left( S_{w}/2\right) !\frac{%
A^{2}}{\Omega ^{S_{w}/2}}\left( 1+\frac{S_{w}}{2}+\frac{k}{\Omega }\right)
+S_{w}!\frac{A^{2}B}{\left( 2\Omega \right) ^{S_{w}+1}}.  \label{Leff2}
\end{equation}%
Then, the Euler-Lagrange equation, $\partial L_{\mathrm{rad}}^{\mathrm{(}%
S_{w}\mathrm{)}}/\partial \left( A^{2}\right) =0$, yields the value of
amplitude $B$, the substitution of which into expression (\ref{NSH}) gives
the instability threshold for the single-color vortex mode:%
\begin{equation}
N_{c}^{(S_{w})}=2^{1+S_{w}}\frac{\left[ \left( S_{w}/2\right) !\right] ^{2}}{%
S_{w}!}\frac{\pi }{\Omega }\left( \Omega +\frac{q}{2}\right) ^{2}.
\label{Nc2}
\end{equation}

The analytical predictions (\ref{Nc})\ and (\ref{Nc2}) produced by the VA
are compared to numerical findings below, see Fig. \ref{figSM}.

\section{Linearized equations for small perturbations}

The stability of the modes under the consideration is the central issue of
the present work. It is addressed via computation of eigenvalues for small
perturbations. In the general case, perturbed solutions are introduced as%
\begin{eqnarray}
u\left( z,\mathbf{r}\right) &=&e^{ikz}\left[ u_{0}(\mathbf{r})+\varepsilon
e^{\lambda z}u_{1}(\mathbf{r})+\varepsilon e^{\lambda ^{\ast }z}\tilde{u}%
_{1}^{\ast }(\mathbf{r})\right] ,  \notag \\
v\left( z,\mathbf{r}\right) &=&e^{ikz}\left[ v_{0}(\mathbf{r})+\varepsilon
e^{\lambda z}v_{1}(\mathbf{r})+\varepsilon e^{\lambda ^{\ast }z}\tilde{v}%
_{1}^{\ast }(\mathbf{r})\right] ,  \label{pert} \\
w\left( z,\mathbf{r}\right) &=&e^{2ikz}\left[ w_{0}(\mathbf{r})+\varepsilon
e^{\lambda z}w_{1}(\mathbf{r})+\varepsilon e^{\lambda ^{\ast }z}\tilde{w}%
_{1}^{\ast }(\mathbf{r})\right] ,  \notag
\end{eqnarray}%
where $k$ is the propagation constant of the stationary solution (assuming
here that $k$ is the same for that both FF components), which is represented
by (complex) functions $u_{0}(\mathbf{r}),~v_{0}(\mathbf{r}),$ $w_{0}(%
\mathbf{r})$, infinitely small $\varepsilon $ is an amplitude of the
perturbation, $\lambda $ is the instability growth rate sought for, which
is, generally, complex too [the instability takes place if there exists, as
usual, at least single $\lambda $ with $\mathrm{Re}(\lambda )>0$], the
asterisk stands for the complex conjugate, and $u_{1}(\mathbf{r}),~\tilde{u}%
_{1}(\mathbf{r}),$ $v_{1}(\mathbf{r}),~\tilde{v}_{1}(\mathbf{r}),$ $w_{1}(%
\mathbf{r}),~\tilde{w}_{1}(\mathbf{r})$ are components of the perturbation
eigenmode, which obey the system of linearized equations:%
\begin{gather}
\left( i\lambda -k\right) u_{1}=-\frac{1}{2}\nabla ^{2}u_{1}+v_{0}^{\ast
}w_{1}+w_{0}\tilde{v}_{1}+\frac{1}{2}\Omega ^{2}r^{2}u_{1}+Qu_{1},  \notag \\
\left( -i\lambda -k\right) \tilde{u}_{1}=-\frac{1}{2}\nabla ^{2}\tilde{u}%
_{1}+v_{0}\tilde{w}_{1}+w_{0}^{\ast }v_{1}+\frac{1}{2}\Omega ^{2}r^{2}\tilde{%
u}_{1}+Q\tilde{u}_{1},  \notag \\
\left( i\lambda -k\right) v_{1}=-\frac{1}{2}\nabla ^{2}v_{1}+u_{0}^{\ast
}w_{1}+w_{0}\tilde{u}_{1}+\frac{1}{2}\Omega ^{2}r^{2}v_{1}-Qv_{1},  \notag \\
\left( -i\lambda -k\right) \tilde{v}_{1}=-\frac{1}{2}\nabla ^{2}\tilde{v}%
_{1}+u_{0}\tilde{w}_{1}+w_{0}^{\ast }u_{1}+\frac{1}{2}\Omega ^{2}r^{2}\tilde{%
v}_{1}-Q\tilde{v}_{1},  \notag \\
\left( 2i\lambda +q-4k\right) w_{1}=-\frac{1}{2}\nabla
^{2}w_{1}+u_{0}v_{1}+v_{0}u_{1}+2\Omega ^{2}r^{2}w_{1},  \notag \\
\left( -2i\lambda +q-4k\right) \tilde{w}_{1}=-\frac{1}{2}\nabla ^{2}\tilde{w}%
_{1}+u_{0}^{\ast }\tilde{v}_{1}+v_{0}^{\ast }\tilde{u}_{1}+2\Omega ^{2}r^{2}%
\tilde{w}_{1}.  \label{6}
\end{gather}

For the unperturbed solution taken as per Eqs. (\ref{uv}) and (\ref{w2}),
it is easy to see that Eqs. (\ref{6}) admit self-consistent perturbation modes
in the following form:%
\begin{eqnarray}
u_{1}(r,\theta ) &=&U_{1}(r)e^{i\left( -S_{u}+p\right) \theta },~\tilde{u}%
_{1}(r,\theta )=\tilde{U}_{1}(r)e^{i\left( S_{u}+p\right) \theta },  \notag
\\
v_{1}(r,\theta ) &=&V_{1}(r)e^{i\left( -S_{v}+p\right) \theta },~\tilde{v}%
_{1}(r,\theta )=\tilde{V}_{1}(r)e^{i\left( S_{v}+p\right) \theta },  \notag
\\
w_{1}(r,\theta ) &=&W_{1}(r)e^{i(-S_{u}-S_{v}+p)\theta },~\tilde{w}%
_{1}(r,\theta )=\tilde{W}_{1}(r)e^{i(S_{u}+S_{v}+p)\theta },  \label{UVW}
\end{eqnarray}%
where $p$ is an arbitrary integer (azimuthal index of the perturbation). The
substitution of expressions (\ref{UVW}) into Eq. (\ref{6}) leads to the
following eigenvalue system which should determine $\lambda $ for each
integer $p$:%
\begin{gather}
\left( i\lambda -k\right) U_{1}=-\frac{1}{2}\left( \frac{d^{2}}{dr^{2}}+%
\frac{1}{r}\frac{d}{dr}-\frac{\left( -S_{u}+p\right) ^{2}}{r^{2}}\right)
U_{1}+V_{0}(r)W_{1}+W_{0}(r)\tilde{V}_{1}  \notag \\
+\frac{1}{2}\Omega ^{2}r^{2}U_{1}+QU_{1},  \notag \\
\left( -i\lambda -k\right) \tilde{U}_{1}=-\frac{1}{2}\left( \frac{d^{2}}{%
dr^{2}}+\frac{1}{r}\frac{d}{dr}-\frac{\left( S_{u}+p\right) ^{2}}{r^{2}}%
\right) \tilde{U}_{1}+V_{0}(r)\tilde{W}_{1}+W_{0}(r)V_{1}  \notag \\
+\frac{1}{2}\Omega ^{2}r^{2}\tilde{U}_{1}+Q\tilde{U}_{1},  \notag \\
\left( i\lambda -k\right) V_{1}=-\frac{1}{2}\left( \frac{d^{2}}{dr^{2}}+%
\frac{1}{r}\frac{d}{dr}-\frac{\left( -S_{v}+p\right) ^{2}}{r^{2}}\right)
V_{1}+U_{0}(r)W_{1}+W_{0}(r)\tilde{U}_{1}  \notag \\
+\frac{1}{2}\Omega ^{2}r^{2}V_{1}-QV_{1},  \notag \\
\left( -i\lambda -k\right) \tilde{V}_{1}=-\frac{1}{2}\left( \frac{d^{2}}{%
dr^{2}}+\frac{1}{r}\frac{d}{dr}-\frac{\left( S_{v}+p\right) ^{2}}{r^{2}}%
\right) \tilde{V}_{1}+U_{0}(r)\tilde{W}_{1}+W_{0}(r)U_{1}  \notag \\
+\frac{1}{2}\Omega ^{2}r^{2}\tilde{V}_{1}-Q\tilde{V}_{1},  \notag \\
\left( 2i\lambda +q-4k\right) W_{1}=-\frac{1}{2}\left( \frac{d^{2}}{dr^{2}}+%
\frac{1}{r}\frac{d}{dr}-\frac{(-S_{u}-S_{v}+p)^{2}}{r^{2}}\right)
W_{1}+U_{0}(r)V_{1}  \notag \\
+V_{0}(r)U_{1}+2\Omega ^{2}r^{2}W_{1},  \notag \\
\left( -2i\lambda +q-4k\right) \tilde{W}_{1}=-\frac{1}{2}\left( \frac{d^{2}}{%
dr^{2}}+\frac{1}{r}\frac{d}{dr}-\frac{(S_{u}+S_{v}+p)^{2}}{r^{2}}\right)
\tilde{W}_{1}  \notag \\
+U_{0}(r)\tilde{V}_{1}+V_{0}(r)\tilde{U}_{1}+2\Omega ^{2}r^{2}\tilde{W}_{1}.
\label{eigen}
\end{gather}%
Solutions $U_{1},\tilde{U}_{1},V_{1},\tilde{V}_{1}$ of Eqs. (\ref{eigen})
must exponentially decay at $r\rightarrow \infty $, and behave as $r^{|p\pm
S_{u,v}|}$ at $r\rightarrow 0$, and $W_{1},\tilde{W}_{1}$ must exponentially
decay too at $r\rightarrow \infty $, and go as $r^{|p\pm \left(
S_{u}+S_{v}\right) |}$ at $r\rightarrow 0$, cf. Eq. (\ref{asympt}).

\section{The stability of the single-color modes}

First, we address the stability of the SH-only vortex, which is given by
Eqs. (\ref{W0}) and (\ref{k}), against small perturbations seeded in the FF
components, which are taken as%
\begin{eqnarray}
u\left( z,\mathbf{r}\right) &=&\varepsilon e^{ikz+\lambda
z}u_{1}(r)e^{ip\theta },  \notag \\
v\left( z,\mathbf{r}\right) &=&\varepsilon e^{ikz+\lambda ^{\ast
}z}v_{1}^{\ast }(r)e^{i\left( S_{w}-p\right) \theta },  \label{P}
\end{eqnarray}%
according to Eqs. (\ref{pert}), (\ref{UVW}). The perturbation
eigenfunctions, $u_{1}(r)$ and $v_{1}(r)$, satisfy the following linearized
equations, which are a particular case of Eq. (\ref{eigen}):%
\begin{gather}
\left( i\lambda -k-Q\right) u_{1}=-\frac{1}{2}\left( \frac{d^{2}}{dr^{2}}+%
\frac{1}{r}\frac{d}{dr}-\frac{P^{2}}{r^{2}}\right) u_{1}  \notag  \label{PP}
\\
+Br^{S_{w}}e^{-\Omega r^{2}}v_{1}+\frac{1}{2}\Omega ^{2}r^{2}u_{1},  \notag
\\
\left( -i\lambda -k+Q\right) v_{1}=-\frac{1}{2}\left( \frac{d^{2}}{dr^{2}}+%
\frac{1}{r}\frac{d}{dr}-\frac{\left( S_{w}-P\right) ^{2}}{r^{2}}\right) v_{1}
\notag \\
+Br^{S_{w}}e^{-\Omega r^{2}}u_{1}+\frac{1}{2}\Omega ^{2}r^{2}v_{1}.  \notag
\end{gather}

The objective is to find, for given $S_{w}$, a critical (minimum) value of
amplitude $B$ in solution (\ref{W0}), and, accordingly, the minimum value of
the integral power (\ref{NSH}), at which Eq. (\ref{PP}) starts to produce
eigenvalues with $\mathrm{Re}(\lambda )\neq 0$, i.e., the single-color mode
becomes unstable against the FF perturbations. Then, we aim to explore the
evolution of unstable modes by means of direct simulations. To address these
problems, we employed numerical techniques and grids similar to those used
in Ref. \cite{hidden} for finding the eigenvalues and running direct
simulations.

\begin{figure}[tbp]
\begin{tabular}{cc}
\includegraphics[width=5.cm]{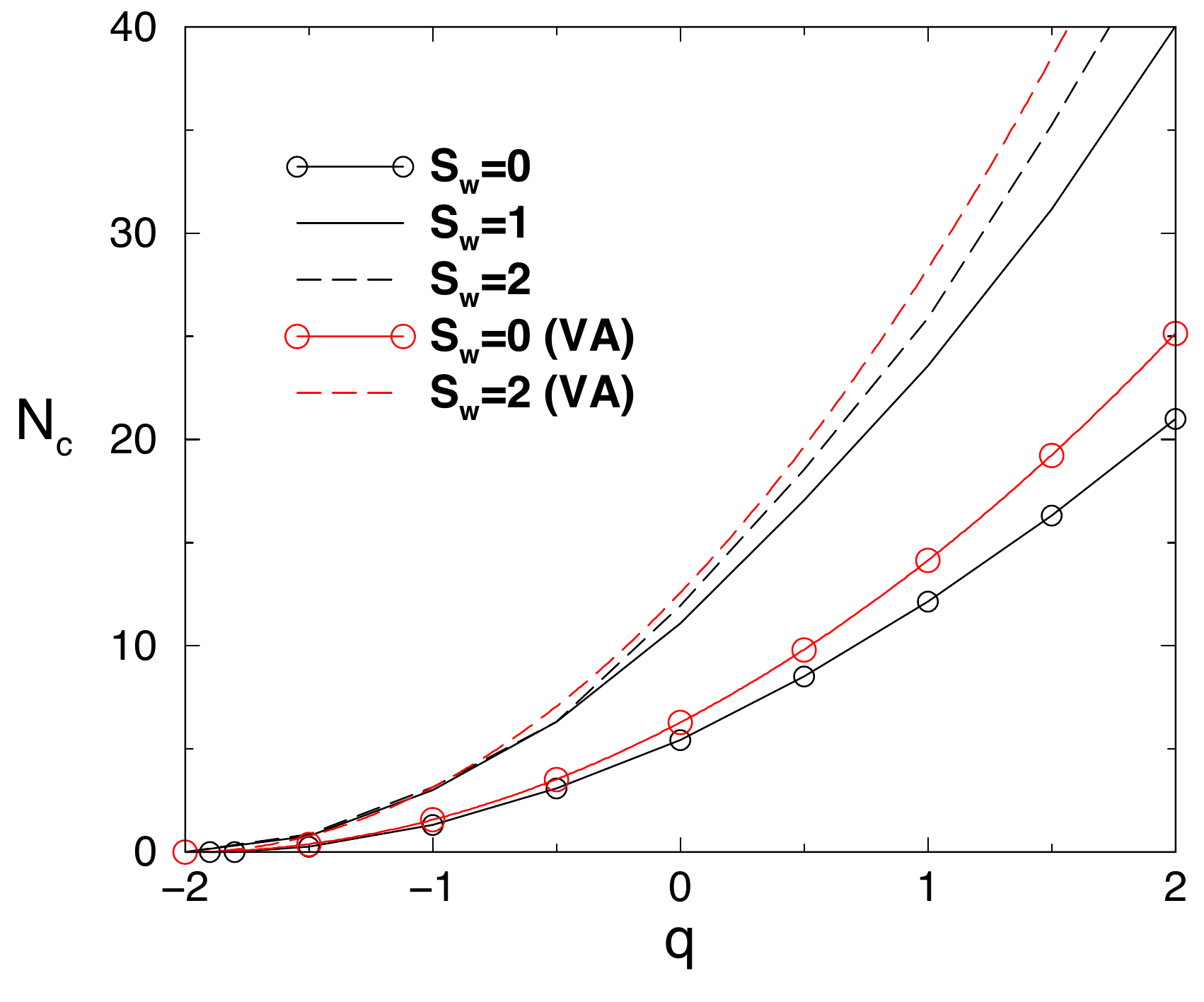} & %
\includegraphics[width=5.cm]{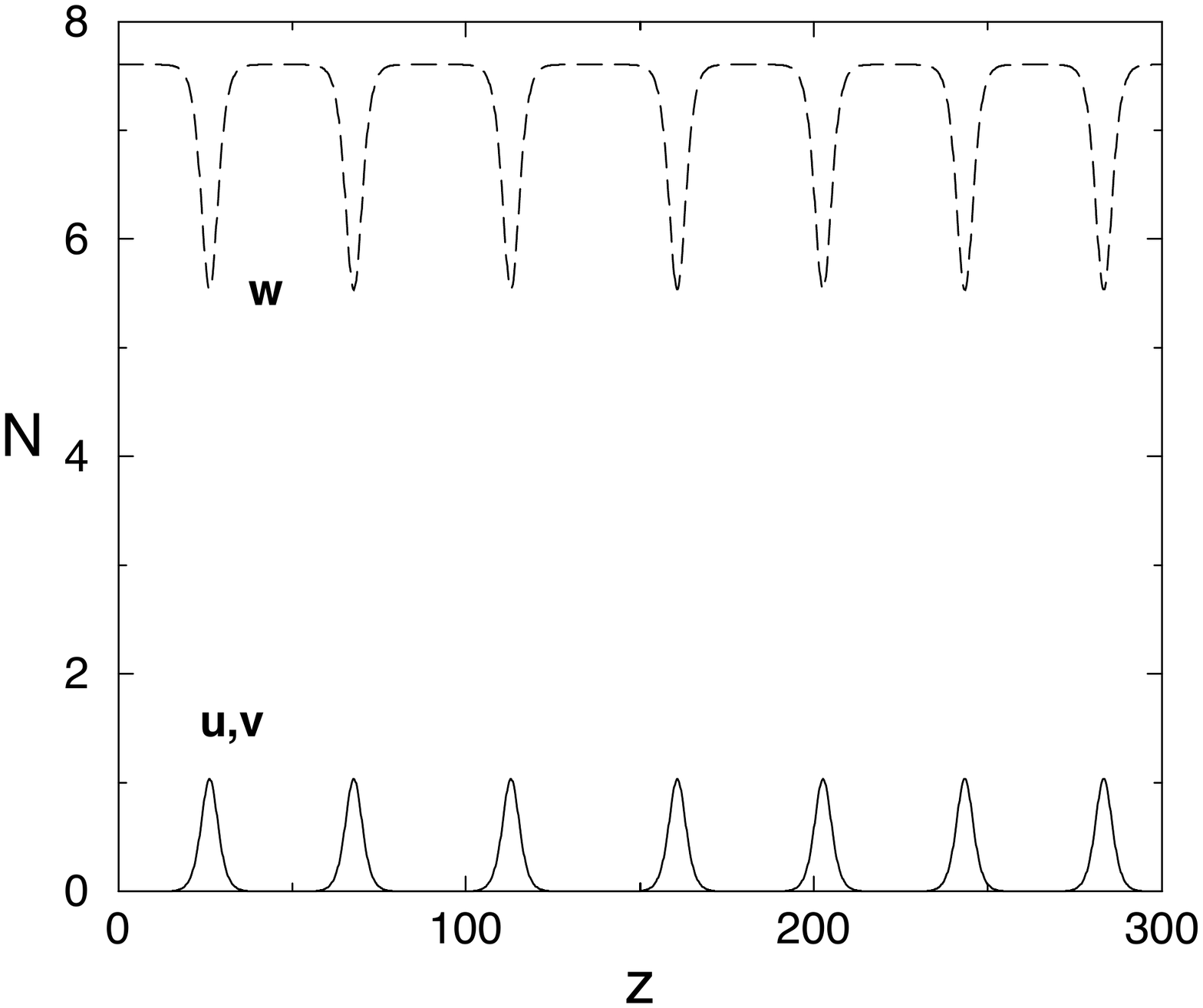} \\
&
\end{tabular}%
\caption{ Left: Numerically found critical powers, $N_{c}$, of the
single-color (SH-only) states, with vorticities $S_{w}=0,1,2$, beyond which
the states become unstable against the excitation of the FF. For $S_{w}=0$
and $2$, the analytical results, produced by the variational approximation
[see Eqs. (\protect\ref{Nc}) with $S=0$ and (\protect\ref{Nc2}) with $%
S_{w}=2 $], are displayed too. Right: The evolution of integral powers of
fields $u,v $ and $w$ for a perturbed unstable state with $S_{w}=0$, $q=0$,
and total power $N=7.6$. In both panels here, and in figures following
below, numerical results are presented for $\Omega =1$ and $Q=0$ (zero
birefringence between the FF components, except for Fig. \protect\ref%
{figm112Q1q0}).}
\label{figSM}
\end{figure}

The results for trapped single-color (SH-only) modes are collected in the
left panel of Fig. \ref{figSM}, while the right panel illustrates the
evolution of an unstable mode with $S_{w}=0$, in terms of the power exchange
between the three components (the total power is conserved, as it should
be). Here and below, numerical results are displayed for the zero
birefringence ($Q=0$, unless it is specified otherwise) and
trapping-potential strength $\Omega =1$.

The left panel of Fig. \ref{figSM} demonstrates that the VA-predicted
instability thresholds (\ref{Nc}) and (\ref{Nc}) approximate their
numerically found counterparts well enough. In particular, the VA exactly
predicts that (for $\Omega =1$ and $Q=0$), the instability threshold
vanishes ($N_{c}=0$) at $q=-2$.

\begin{figure}[th]
\begin{tabular}{cccc}
\includegraphics[scale=0.17]{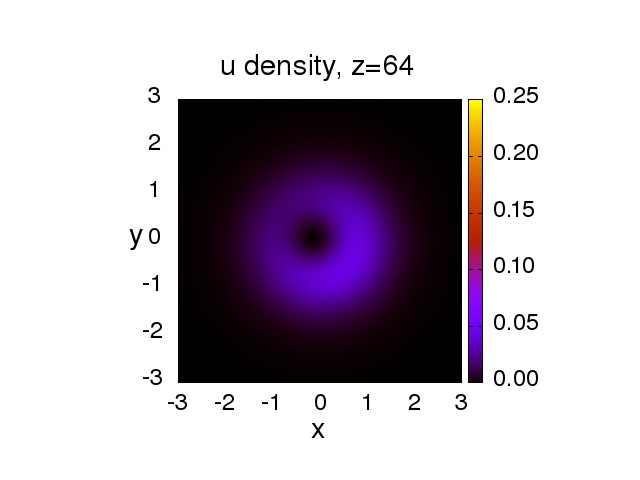} & %
\includegraphics[scale=0.17]{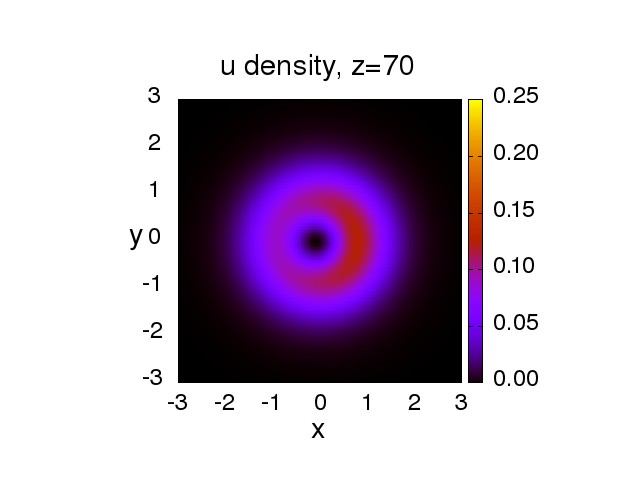} & %
\includegraphics[scale=0.17]{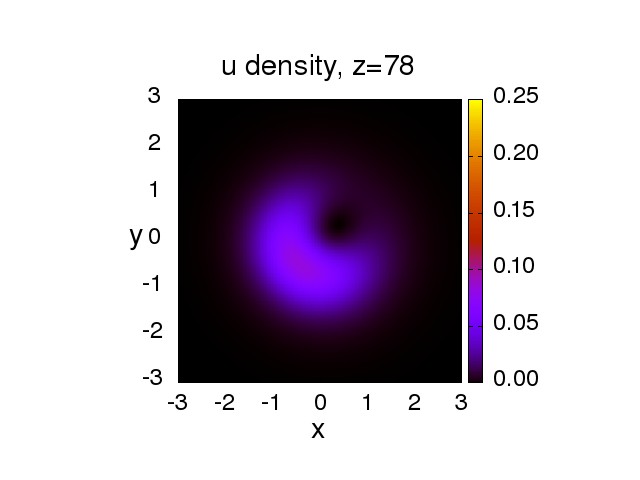} & %
\includegraphics[scale=0.17]{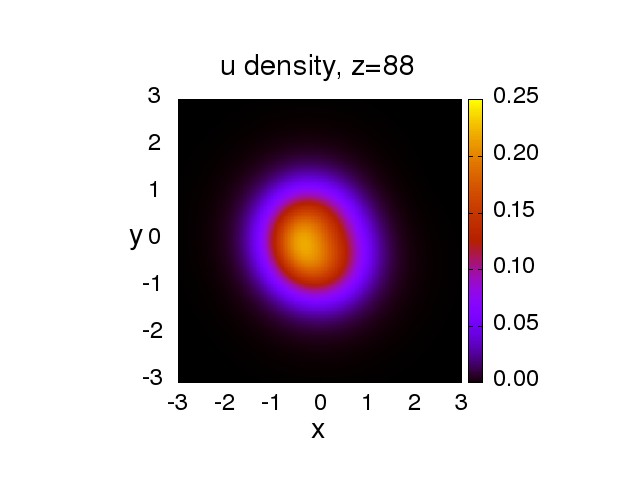} \\
\includegraphics[scale=0.17]{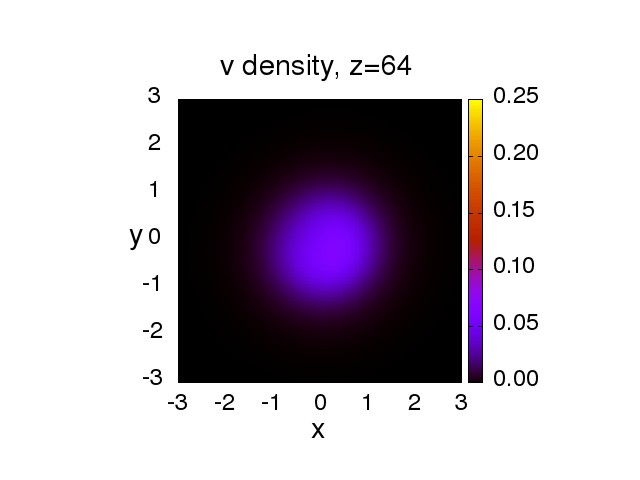} & %
\includegraphics[scale=0.17]{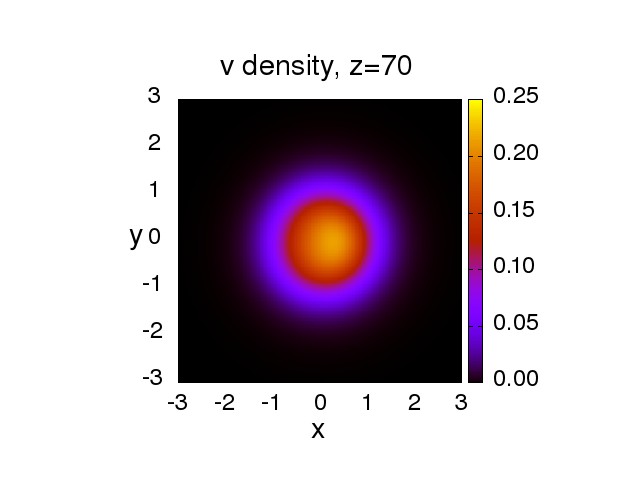} & %
\includegraphics[scale=0.17]{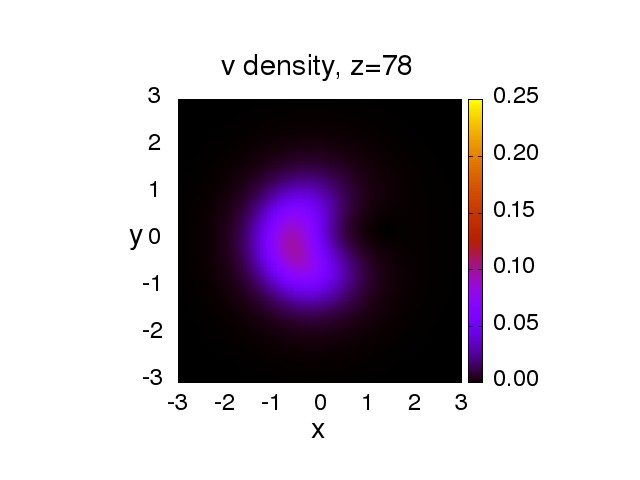} & %
\includegraphics[scale=0.17]{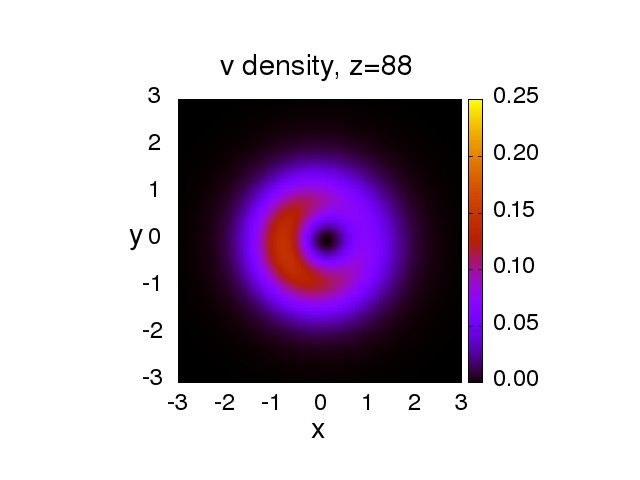} \\
\includegraphics[scale=0.17]{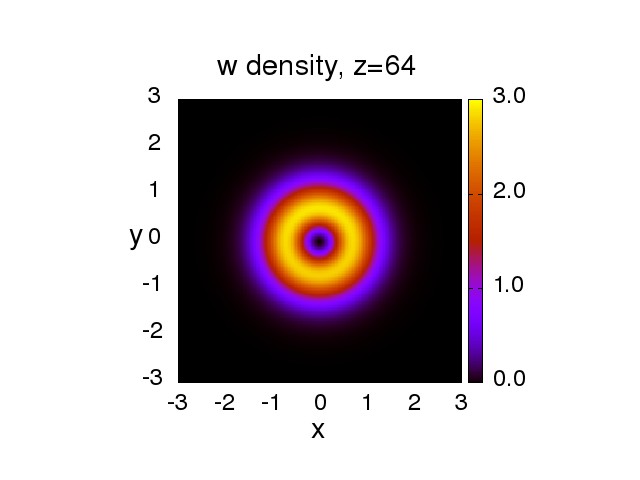} & %
\includegraphics[scale=0.17]{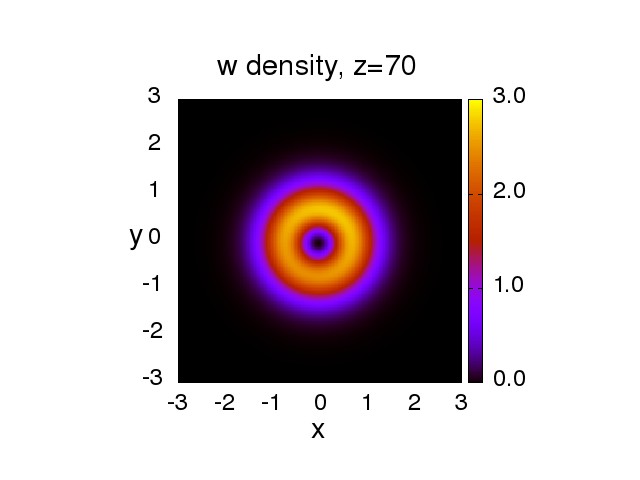} & %
\includegraphics[scale=0.17]{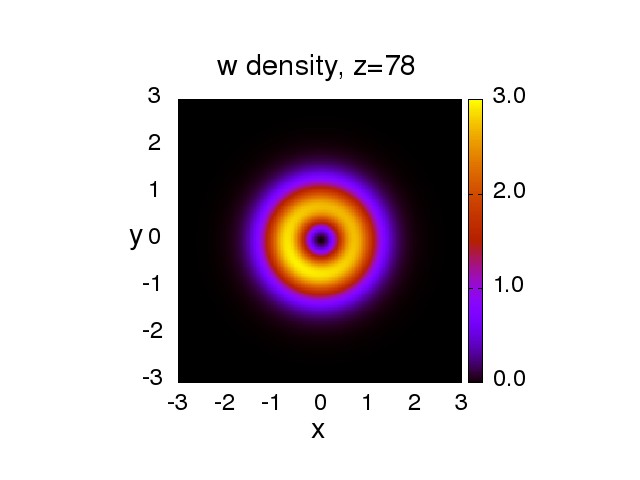} & %
\includegraphics[scale=0.17]{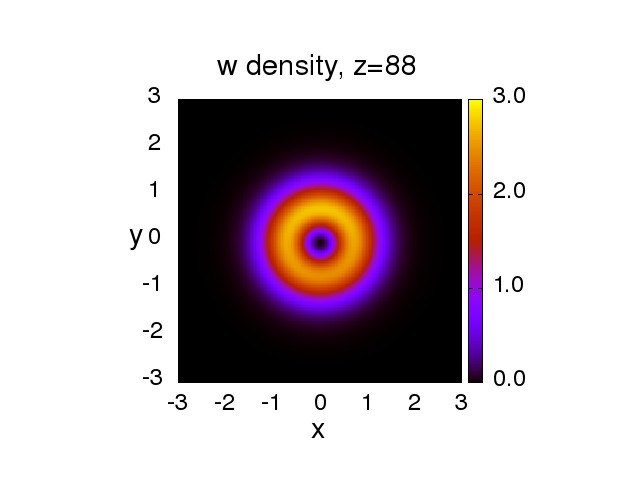} \\
&  &  &
\end{tabular}%
\caption{(Color online) Development of the instability of the single-color
(SH) mode (\protect\ref{W0}), for $S_{w}=1$, $Q=0$, $\Omega =1$, $B=2$, $N=4%
\protect\pi $. Initial random perturbations in the FF components, $u$ and $v$%
, are loaded with an amplitude $<10^{-4}$. The top, middle, and bottom
panels display the evolution of densities of the $u$, $v$ (FF) and $w$ (SH)
fields, respectively.}
\label{fig_SM_evolution}
\end{figure}

\begin{figure}[th]
\begin{tabular}{cc}
\includegraphics[scale=0.4]{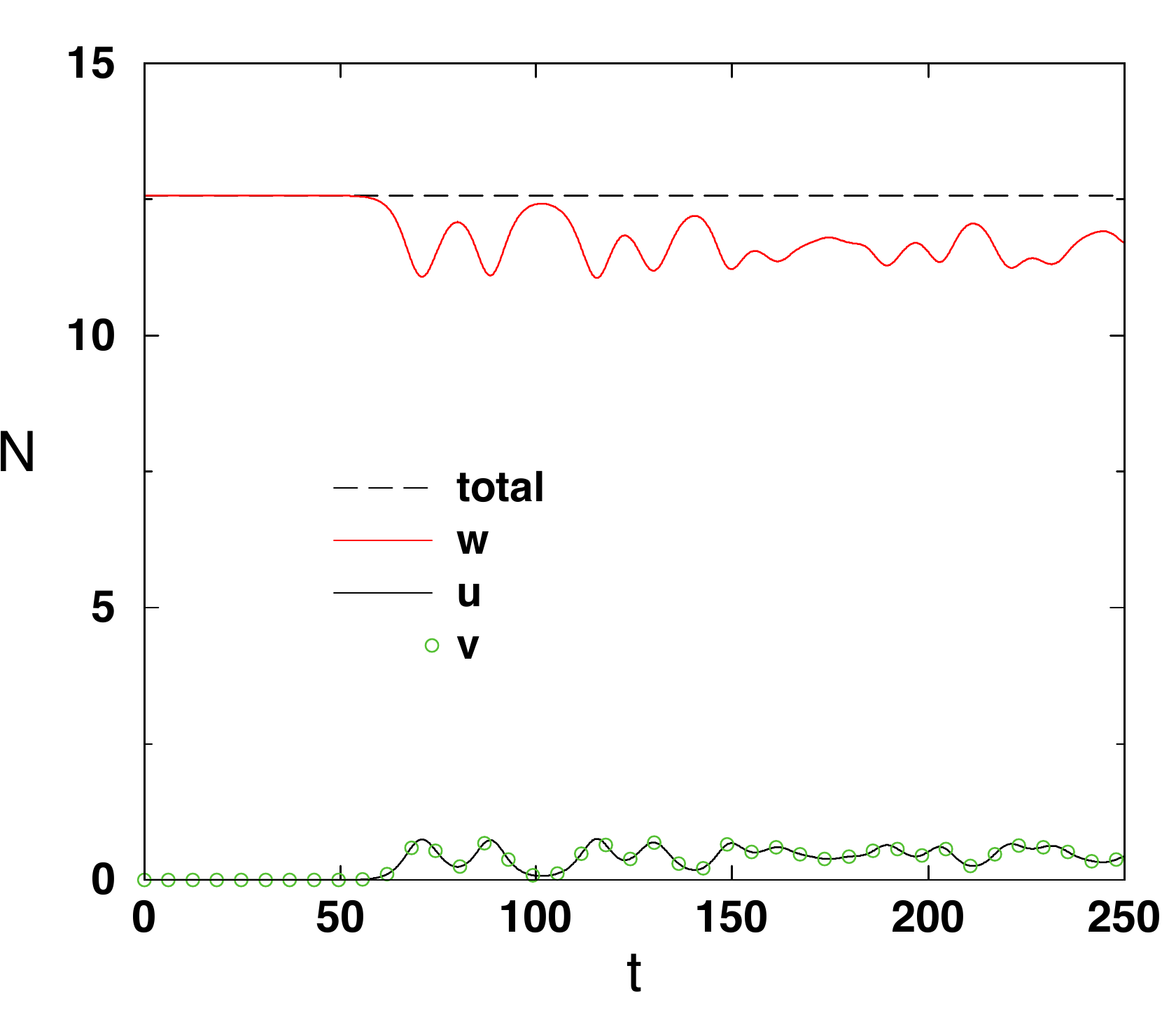} & %
\includegraphics[scale=0.4]{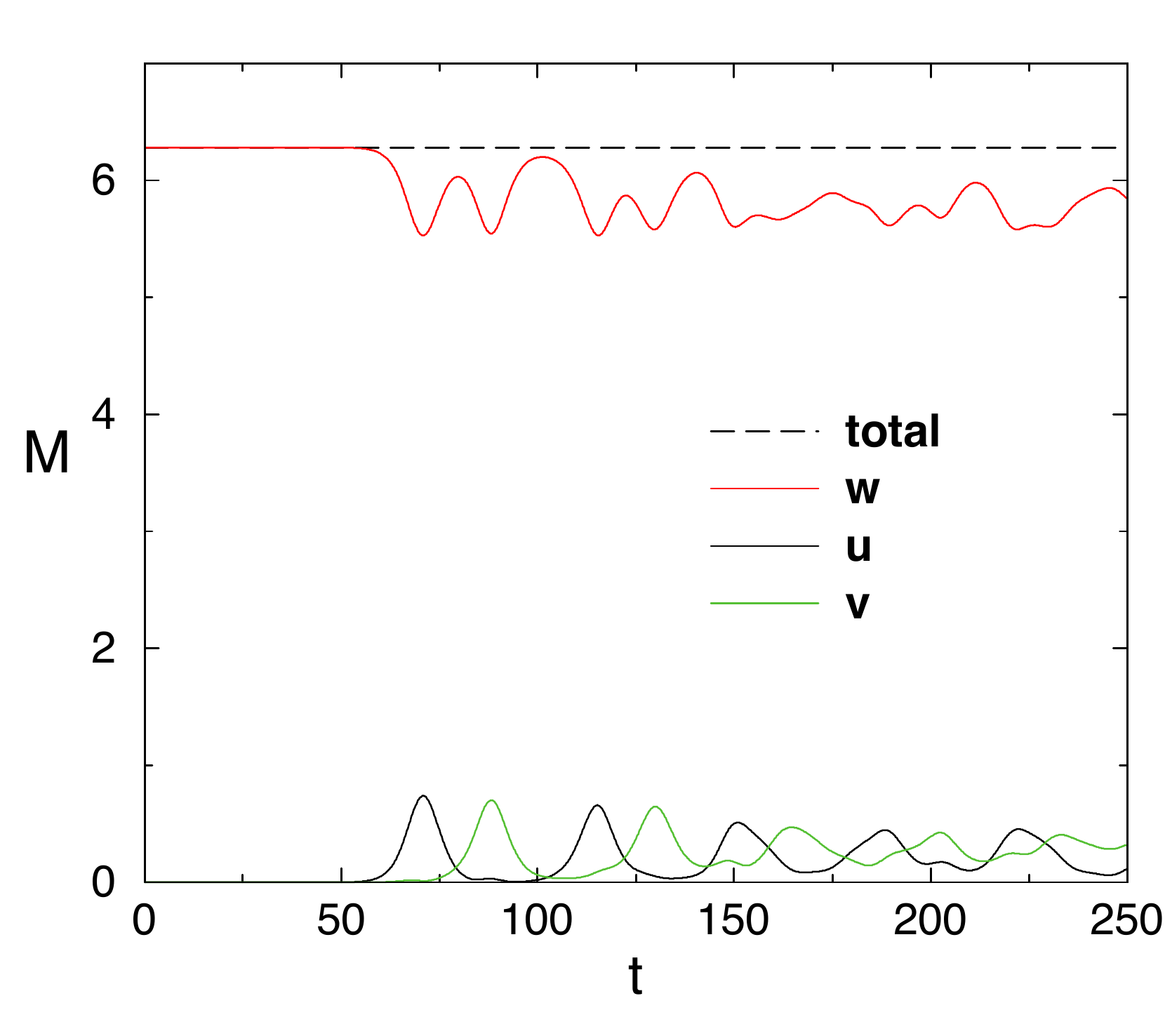} \\
&
\end{tabular}%
\caption{Overall characteristics of the instability displayed in Fig.
\protect\ref{fig_SM_evolution}. Left: The evolution of the integral powers
of the FF components, $u$ and $v$. Right: The integral angular momentum of
each component, see Eq. (\protect\ref{M}). }
\label{fig_SM_evol_normLz}
\end{figure}

It is interesting to consider the (in)stability of the single-color trapped
vortices with $S_{w}=1$ and other odd values of the vorticity, as the
respective perturbations in the two FF components cannot be arranged
symmetrically, unlike those considered above for even values of $S_{w}$. The
threshold value of the norm for $S_{w}=1$ is shown in the left panel of Fig. %
\ref{figSM}, and the development of the respective instability is
illustrated by Figs. (\ref{fig_SM_evolution}) and (\ref{fig_SM_evol_normLz}%
). It can be concluded that the instability does not destroy the vorticity
(topological charge) of the SH component, while the SH and FF waves exchange
their angular momenta, keeping virtually equal integral powers.

\section{Numerical results for three-wave vortices}

\subsection{Hidden-vorticity (HV) modes with $S_{u,v,w}=\left(
+1,-1,0\right) $}

\begin{figure}[th]
\includegraphics[width=5.5cm,clip]{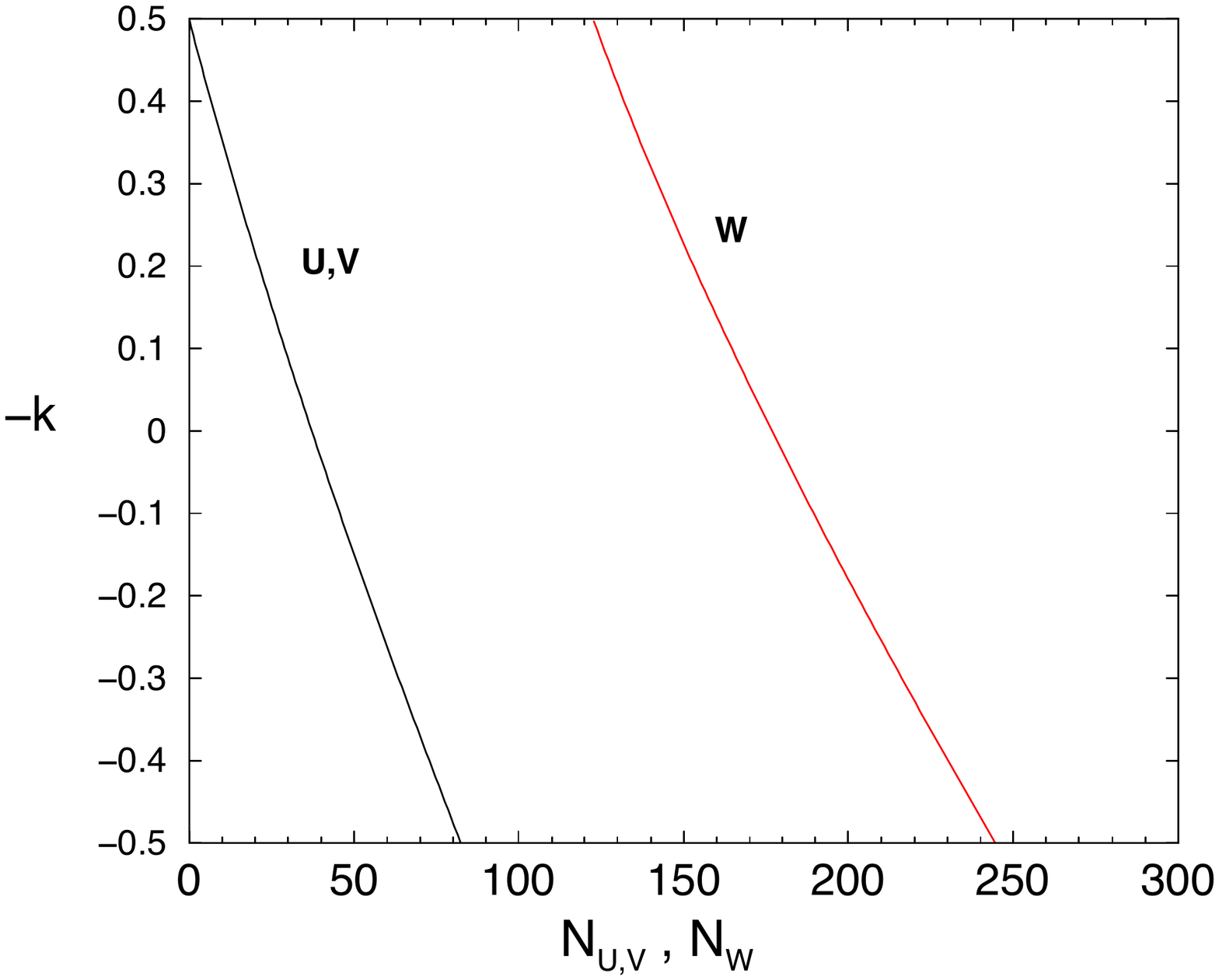} %
\includegraphics[width=5cm,clip]{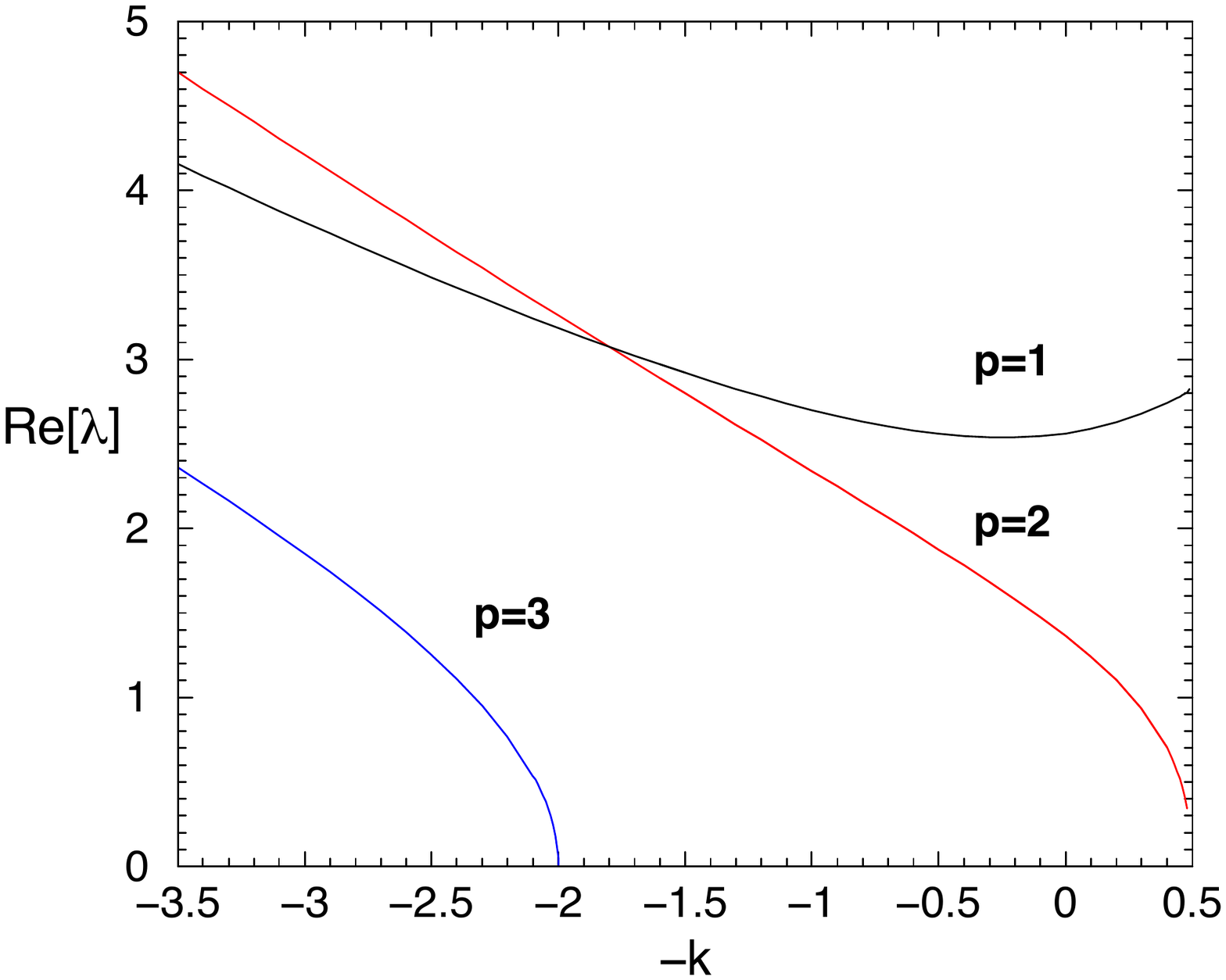}
\caption{(Color online) Left: The propagation constant vs. the total power
for the hidden-vorticity family with vorticities $S_{u,v,w}=\left(
+1,-1,0\right) $, $\Omega =1$, $Q=q=0$. Right: Stability eigenvalues
produced by linearized equations (\protect\ref{eigen}). The corresponding
values of the perturbation azimuthal index $p$ are attached to the curves. }
\label{fig1}
\end{figure}

Families of HV states, defined as per Eqs. (\ref{uv}) and (\ref{w2}), are
characterized by dependences between the propagation constant and integral
powers of the two FF components and the SH one, see Eq. (\ref{N}). Typical
examples of such dependences, obtained from a numerical solution of Eqs. (%
\ref{U})-(\ref{W}), are shown in the left panels of Figs. \ref{fig1} and \ref%
{fig1b}, respectively, for positive ($q\geq 0$) and negative ($q<0$) values
of the phase mismatch.
\begin{figure}[th]
\includegraphics[width=5.5cm,clip]{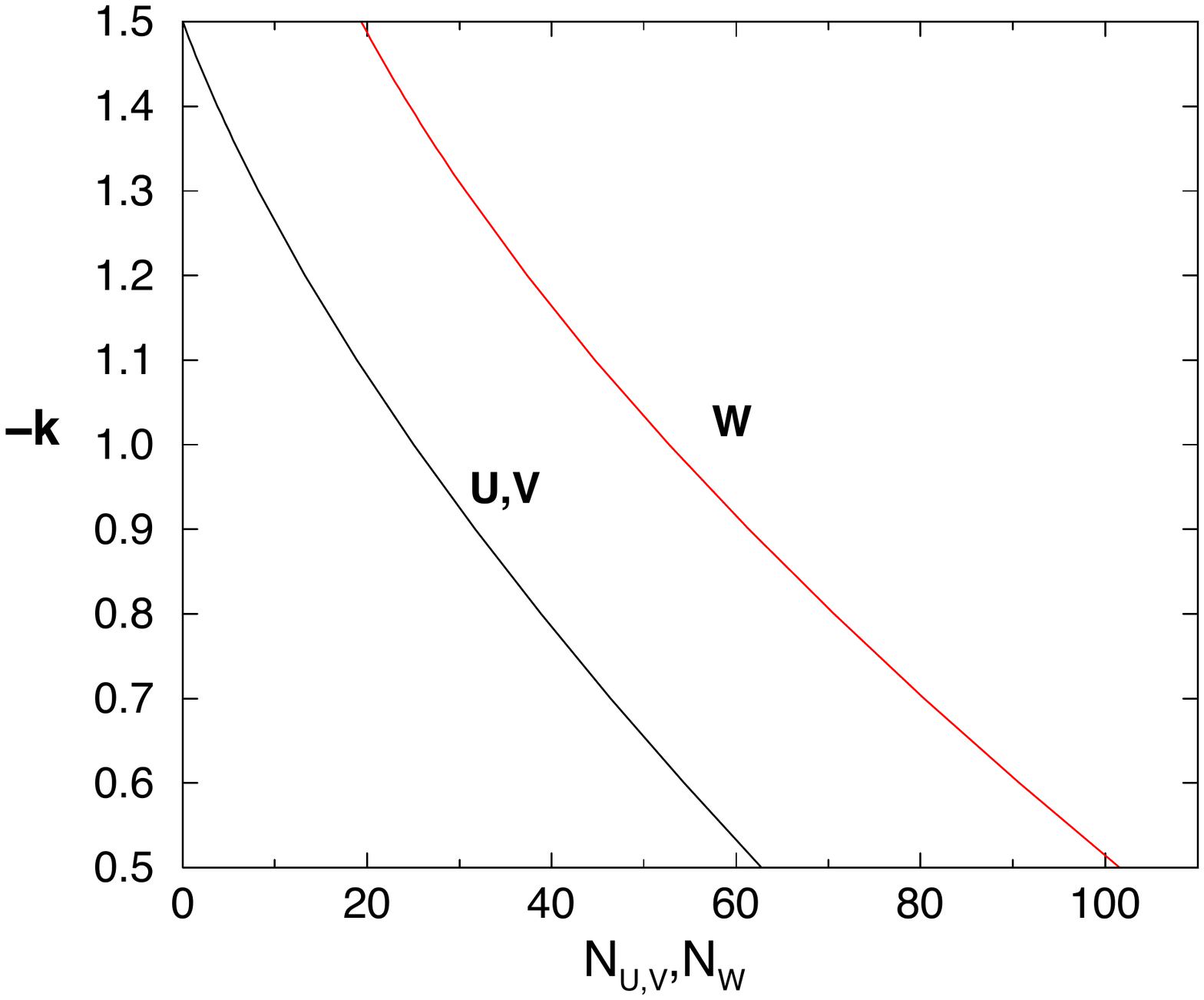} %
\includegraphics[width=5cm,clip]{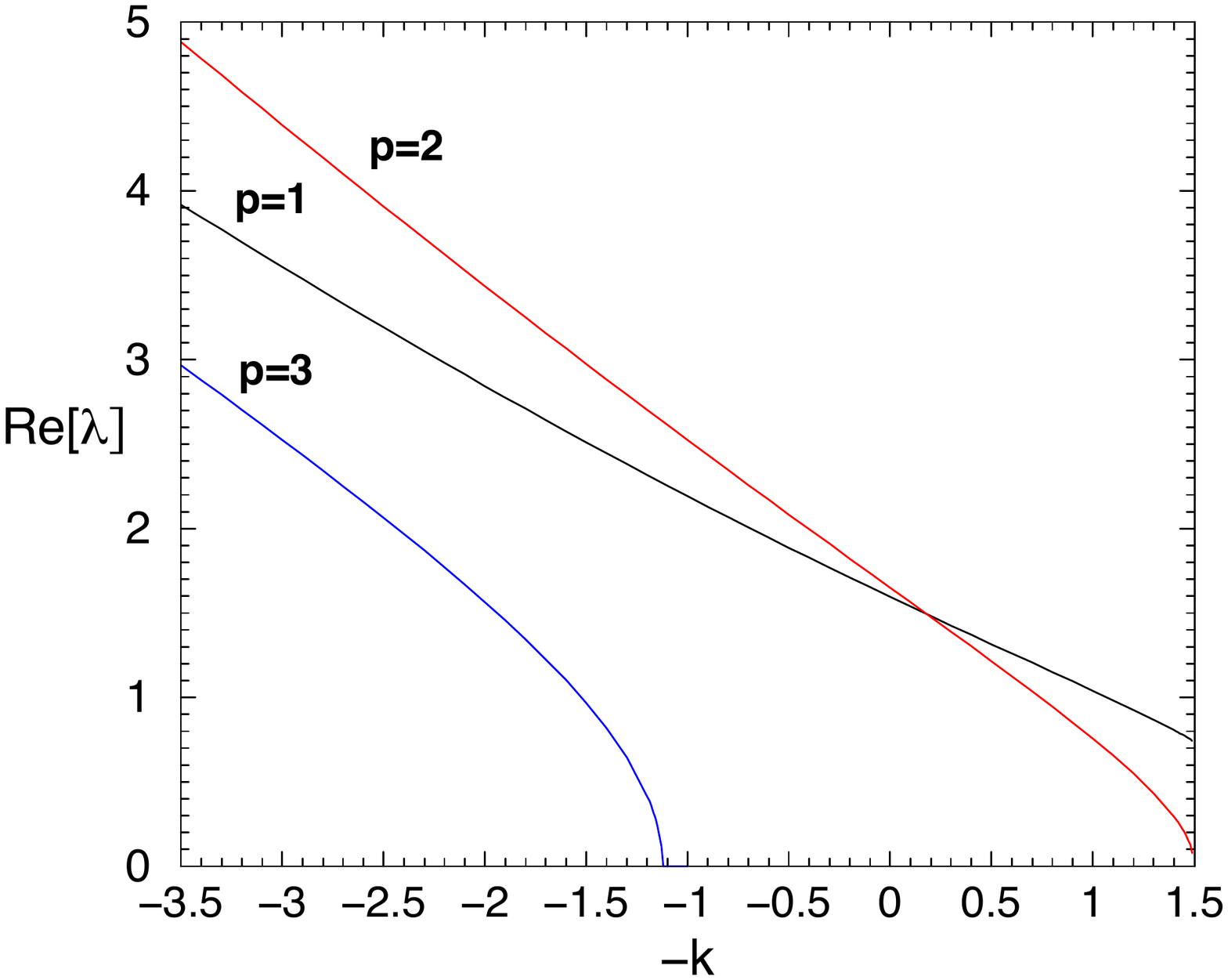}
\caption{(Color online) The same as in Fig. \protect\ref{fig1}, but for $%
q=-4 $.}
\label{fig1b}
\end{figure}

\begin{figure}[th]
\begin{tabular}{cccc}
\includegraphics[scale=0.17]{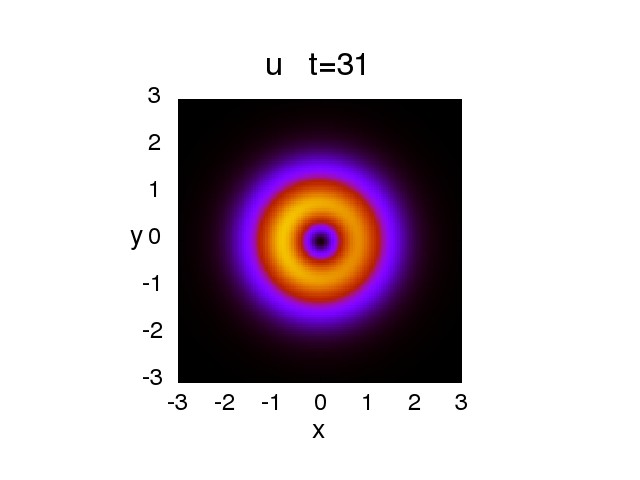} & %
\includegraphics[scale=0.17]{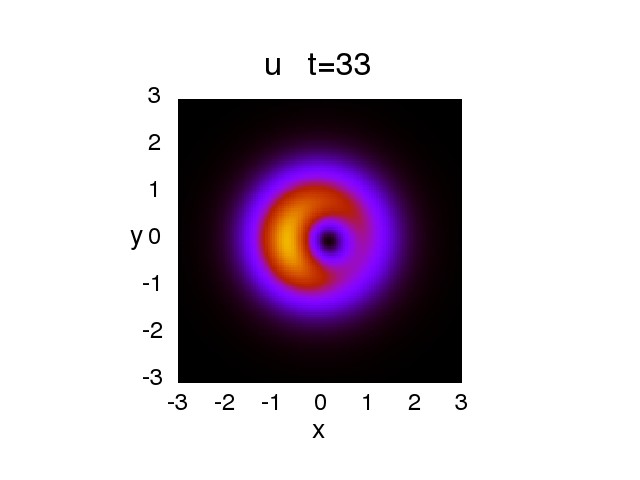} & %
\includegraphics[scale=0.17]{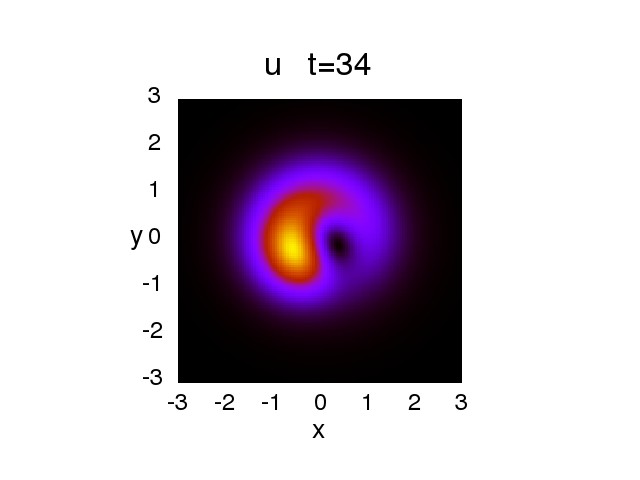} & %
\includegraphics[scale=0.17]{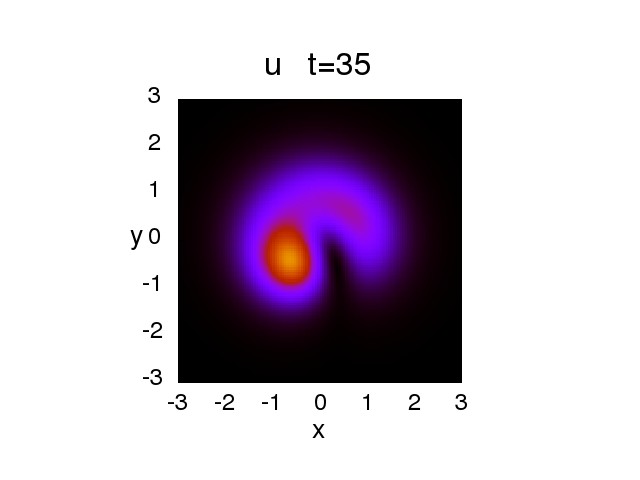} \\
&  &  &
\end{tabular}%
\caption{(Color online) The evolution of the two components, $u$, of an
unstable hidden-vorticity mode with $q=-4$, norms $N_{u,v}=25.1$, $%
N_{w}=52.8 $ (the total power is $N=103$) and $k=-1$. Density distributions
are displayed for values of the propagation distance indicated above the
respective panels.}
\label{fig4b}
\end{figure}

\begin{figure}[th]
\begin{tabular}{cccc}
\includegraphics[scale=0.17]{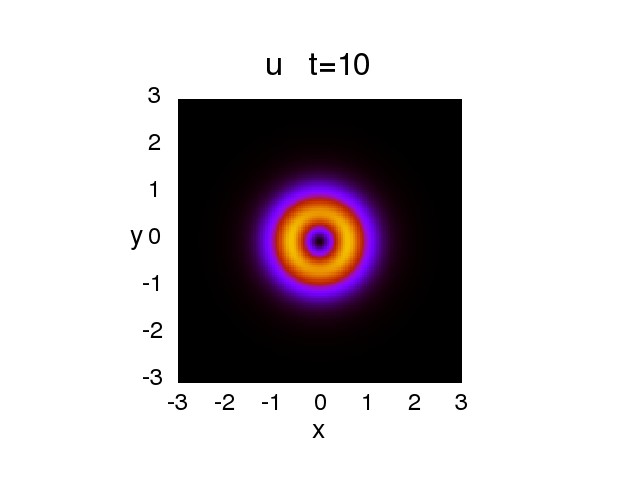} & %
\includegraphics[scale=0.17]{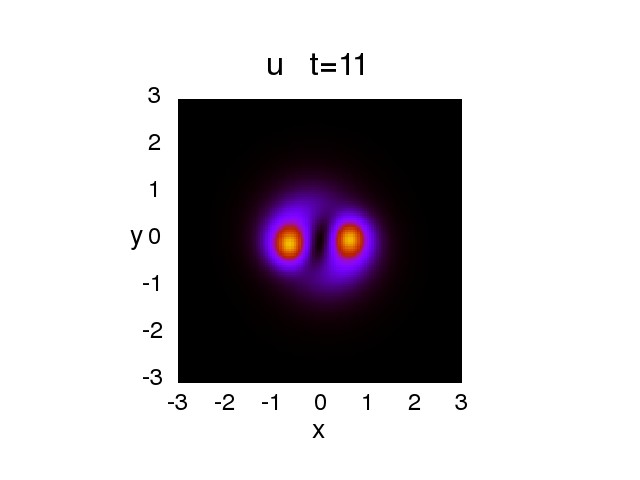} & %
\includegraphics[scale=0.17]{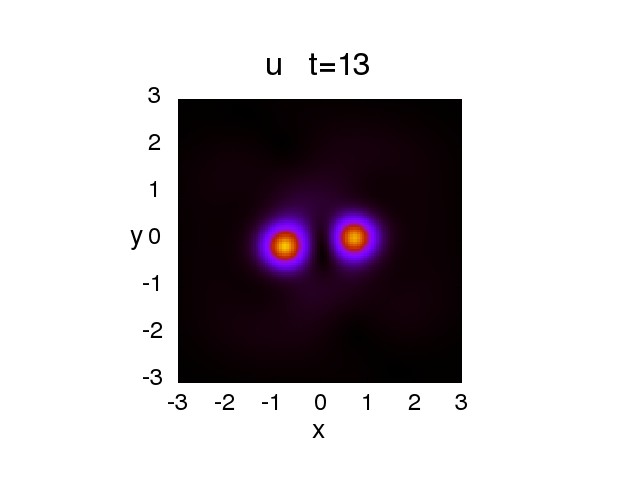} & %
\includegraphics[scale=0.17]{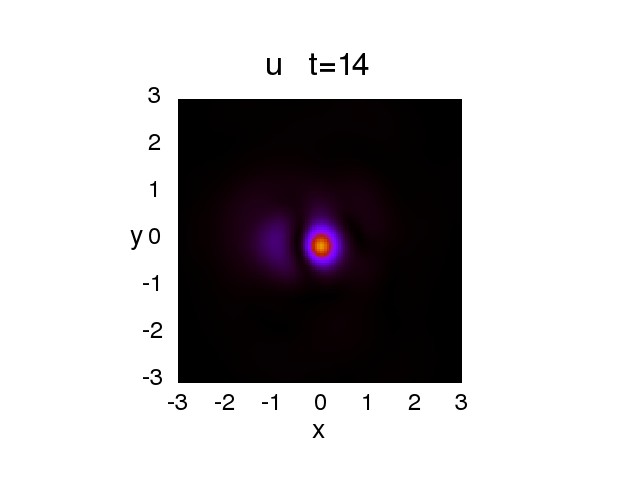} \\
&  &  &
\end{tabular}%
\caption{(Color online) The same as in Fig. \protect\ref{fig4b}, but for $%
q=-4$ and $N_{u,v}=340$, $N_{w}=480$, the total power is $N=1160$, and $k=2$%
. }
\label{fig4c}
\end{figure}

The computation of the stability eigenvalues by means of Eq. (\ref{eigen})
demonstrates, in the right panels of Figs. \ref{fig1} and \ref{fig1b}, that
all the HV modes are subject to instability against small perturbations with
various values of azimuthal index $p$. Direct simulations reveal two
different scenarios of the instability development, which are displayed in
Figs. \ref{fig4b} and \ref{fig4c}. In the former case, the central core is
spontaneously expelled from the vortex ring, which tends to transform itself
into a fundamental mode. In the latter case, the vortex ring splits into two
fragments, which is a typical outcome of the instability development of
vortices in models with the cubic nonlinearity \cite{cubic-in-trap3}-\cite%
{cubic-in-trap6}.

\subsection{Semi-vortices with $S_{u,v,w}=\left( 1,0,1\right) $}

\begin{figure}[th]
\includegraphics[width=5.5cm,clip]{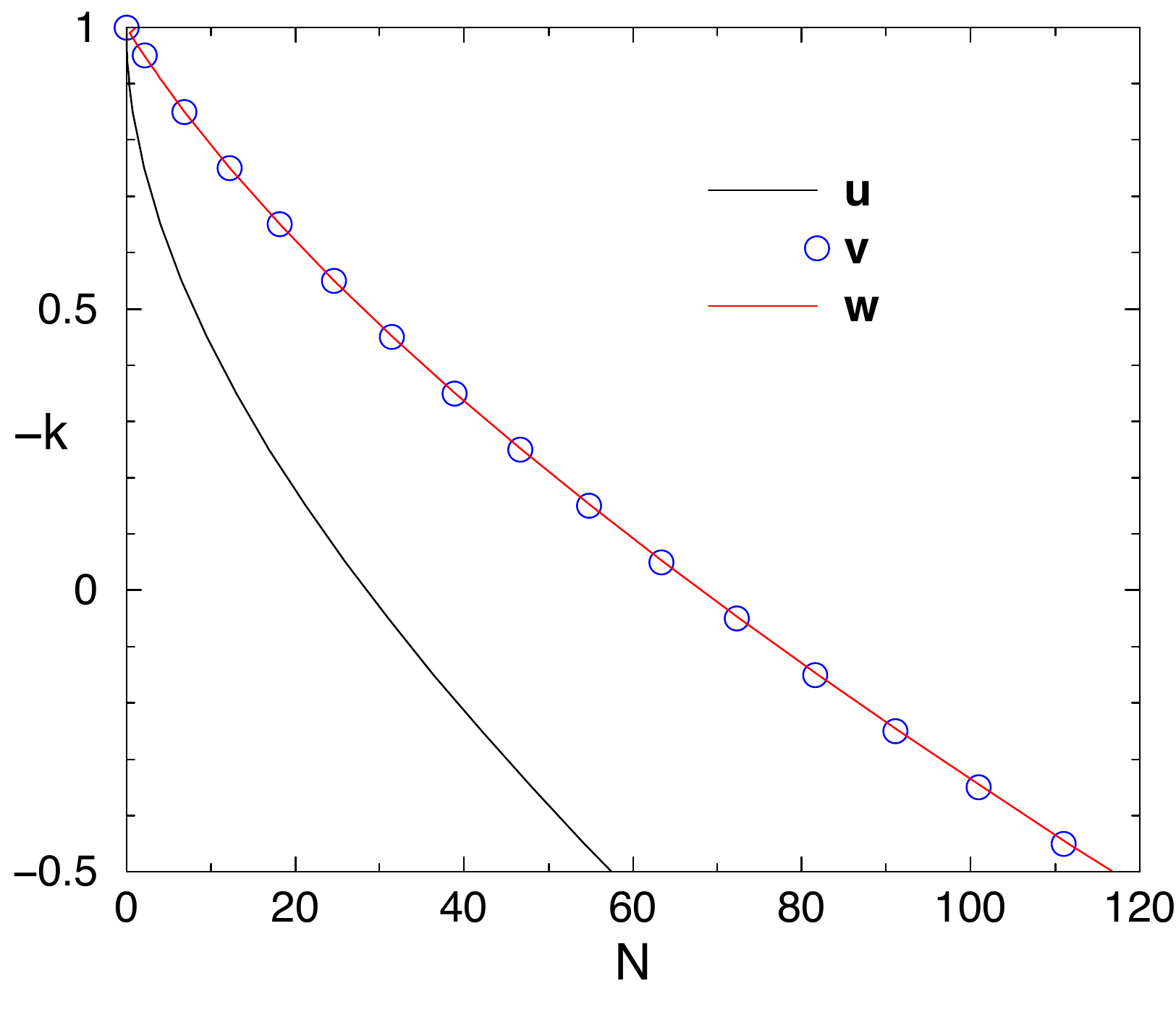} %
\includegraphics[width=5cm,clip]{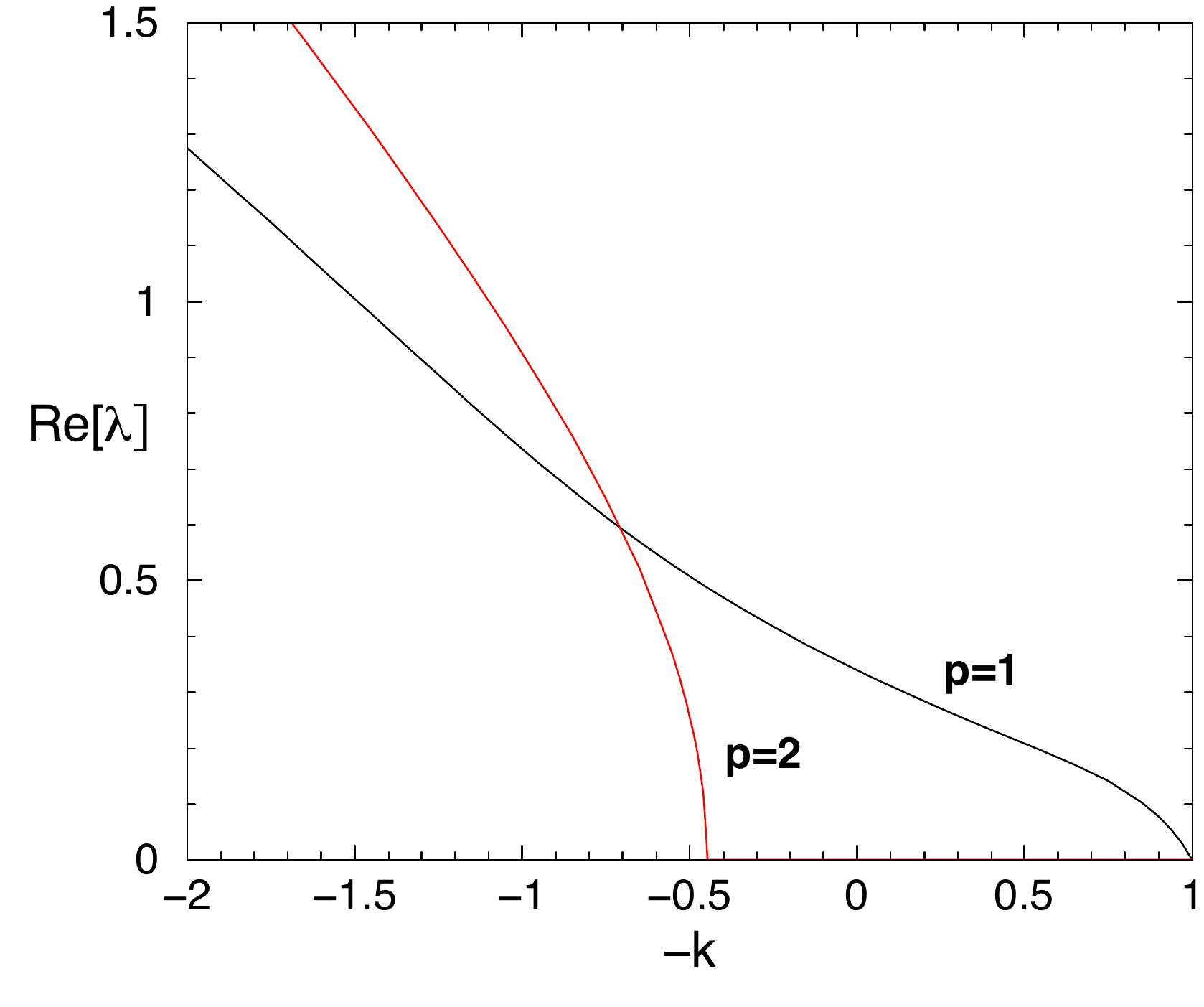}
\caption{(Color online) Left: Relations between the propagation constant and
integral powers of the three fields for semi-vortices of the type of $%
S_{u,v,w}=\left( 1,0,1 \right) $, with $\Omega =1$, $Q=0,$ $q=0$. Right:
The largest real parts of stability eigenvalues for the semi-vortices, as
obtained from the numerical solution of Eq. (\protect\ref{eigen}). Integers $%
p$ are azimuthal indices of the destabilizing perturbation eigenmodes.}
\label{figm101}
\end{figure}

For families of semi-vortex solutions, a numerical solution of Eq. (\ref%
{1-0-1}) produces dependences between the propagation constant and integral
powers of the three components, a typical example of which is displayed in
Fig. \ref{figm101}. It is concluded that these families are completely
unstable too.

\begin{figure}[th]
\begin{tabular}{cccc}
\includegraphics[scale=0.17]{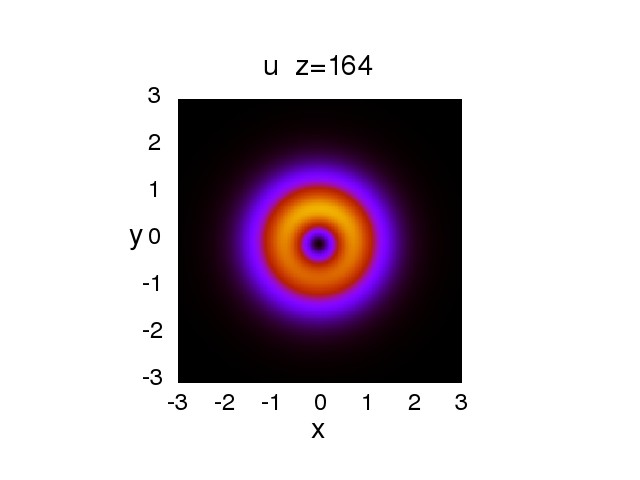} & %
\includegraphics[scale=0.17]{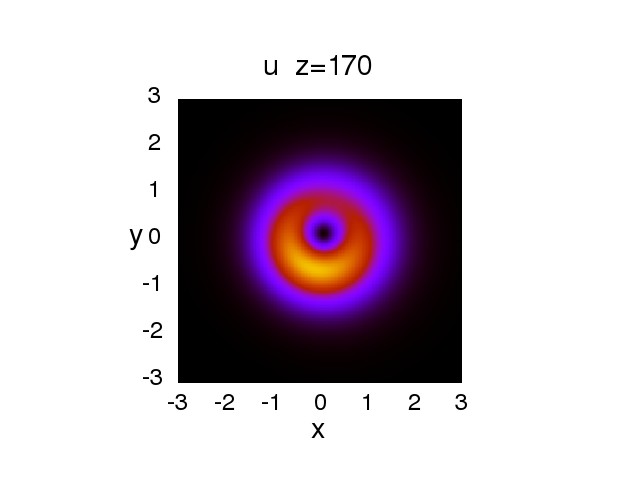} & %
\includegraphics[scale=0.17]{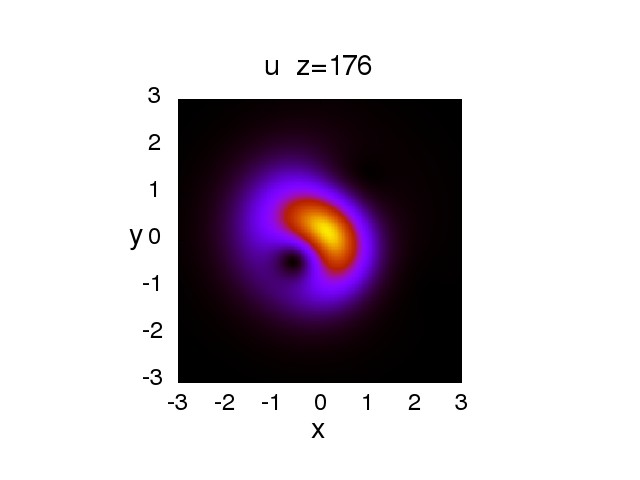} & %
\includegraphics[scale=0.17]{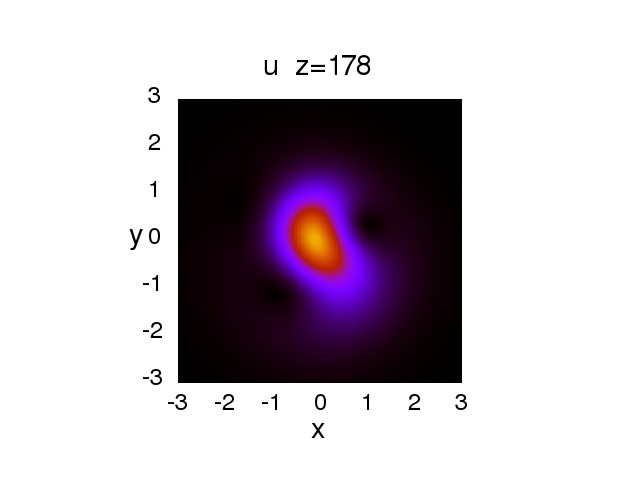} \\
\includegraphics[scale=0.17]{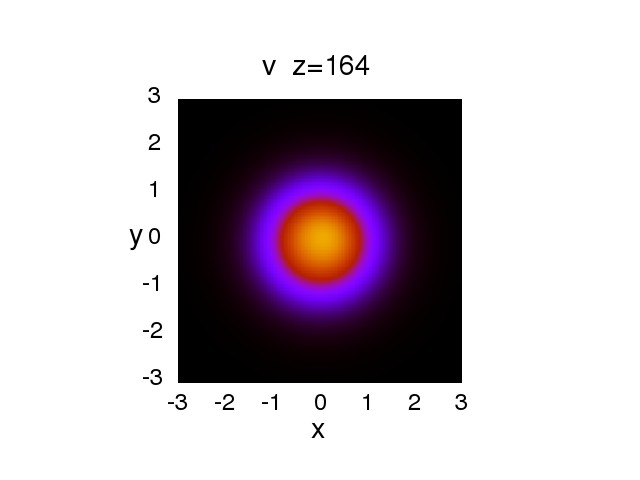} & %
\includegraphics[scale=0.17]{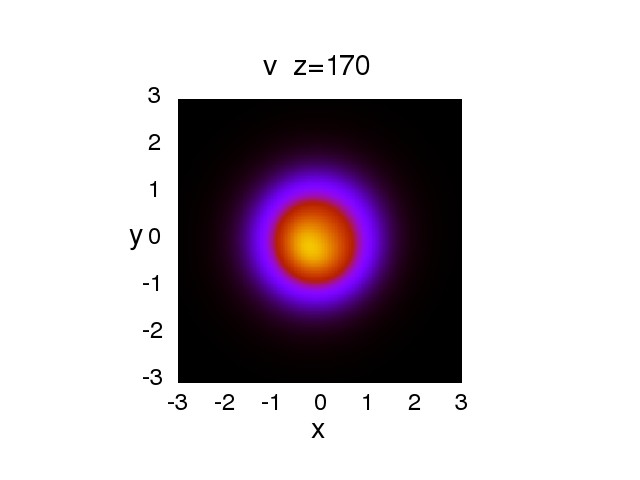} & %
\includegraphics[scale=0.17]{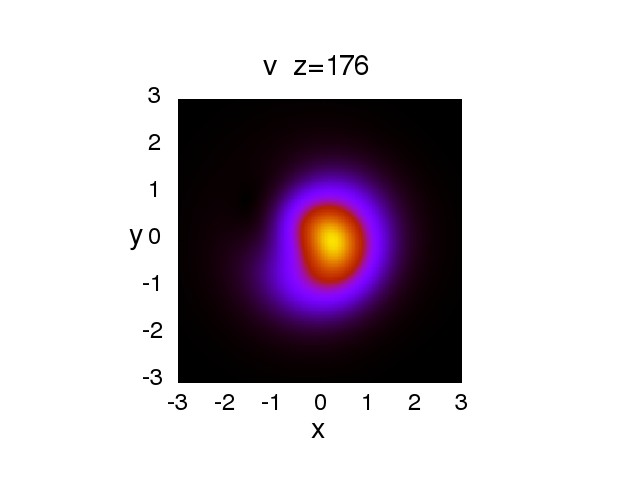} & %
\includegraphics[scale=0.17]{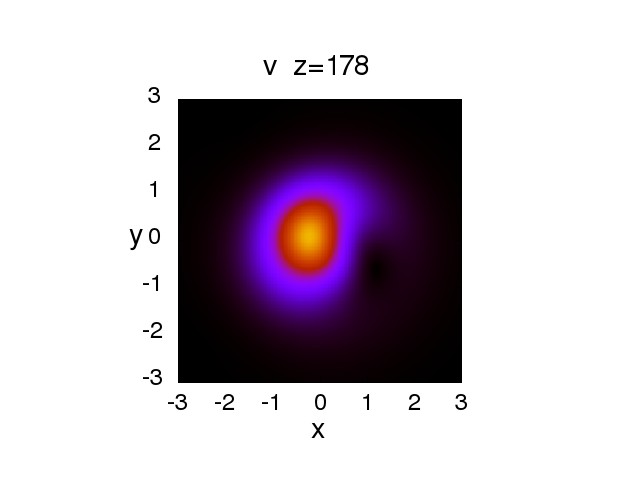} \\
\includegraphics[scale=0.17]{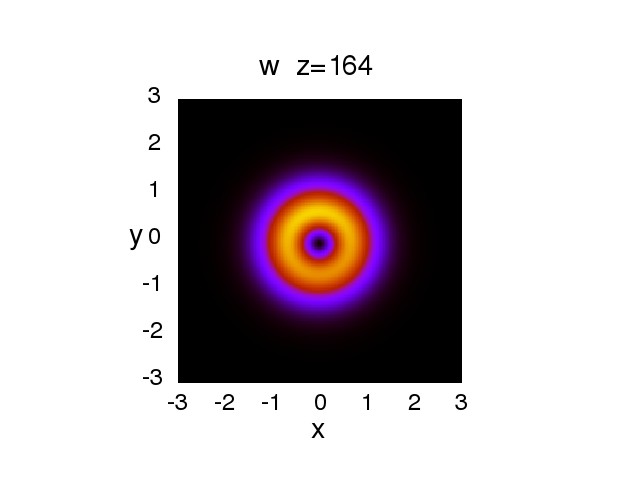} & %
\includegraphics[scale=0.17]{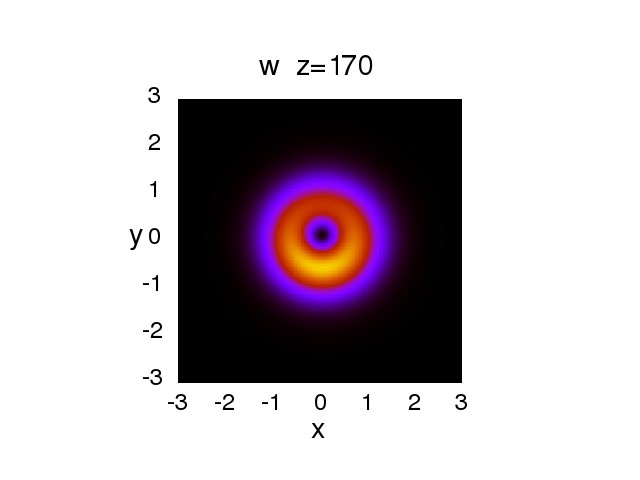} & %
\includegraphics[scale=0.17]{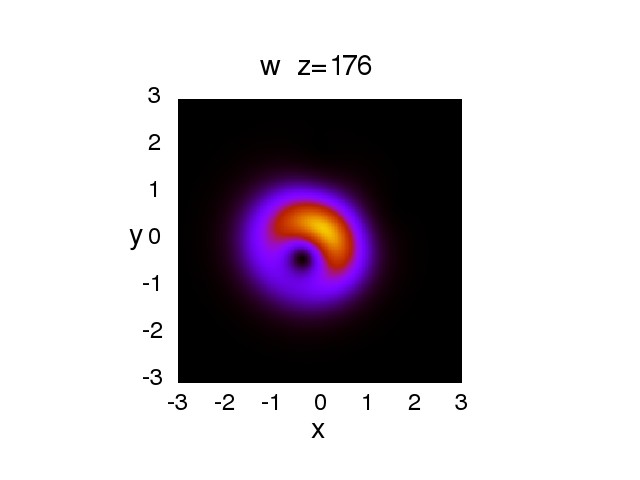} & %
\includegraphics[scale=0.17]{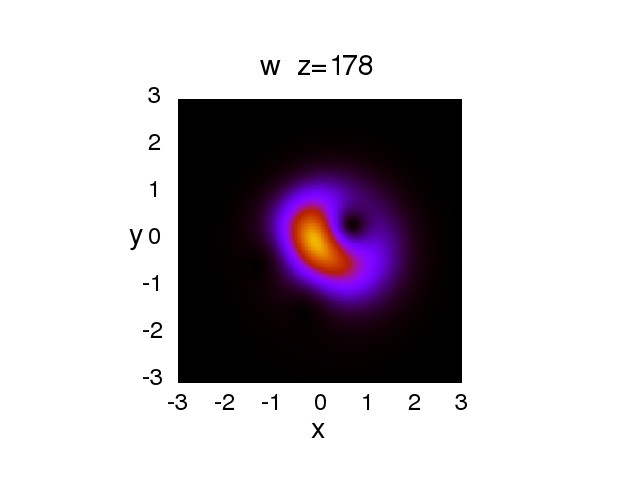} \\
&  &  &
\end{tabular}%
\caption{(Color online) The evolution of the densities of the $u,v,w$ fields
in an unstable semi-vortex corresponding to $k=-0.5$ in Fig.~\protect\ref%
{figm101}. }
\label{figm101_mu0.5}
\end{figure}

Simulations of the instability development of the semi-vortices reveal two
basic outcomes which initially look similar to those shown above for the HV
modes, cf. Figs. \ref{fig4b} and \ref{fig4c}, i.e., expulsion of the central
core from the vortex ring, as shown in Fig. \ref{figm101_mu0.5}, or
splitting of the vortex into a set of two fragments, see Fig. \ref%
{figm101_mu-1}.

\begin{figure}[th]
\begin{tabular}{cccc}
\includegraphics[scale=0.17]{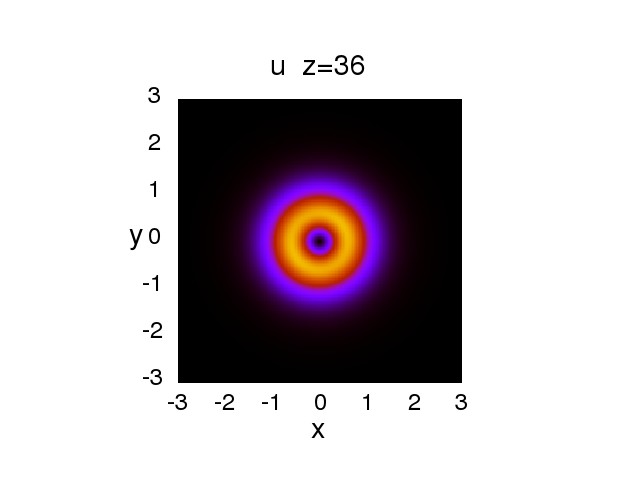} & %
\includegraphics[scale=0.17]{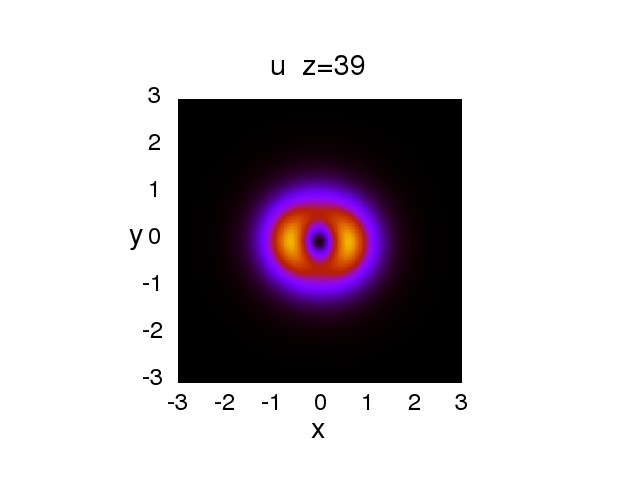} & %
\includegraphics[scale=0.17]{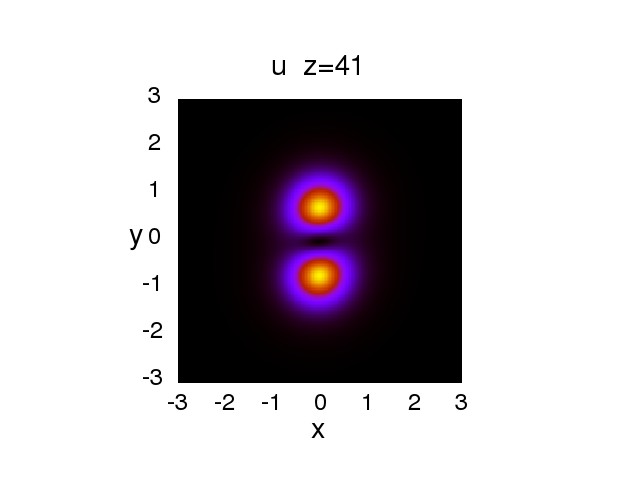} & %
\includegraphics[scale=0.17]{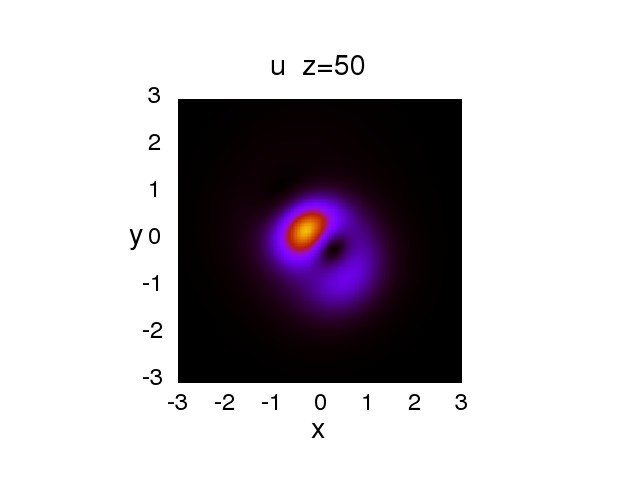} \\
\includegraphics[scale=0.17]{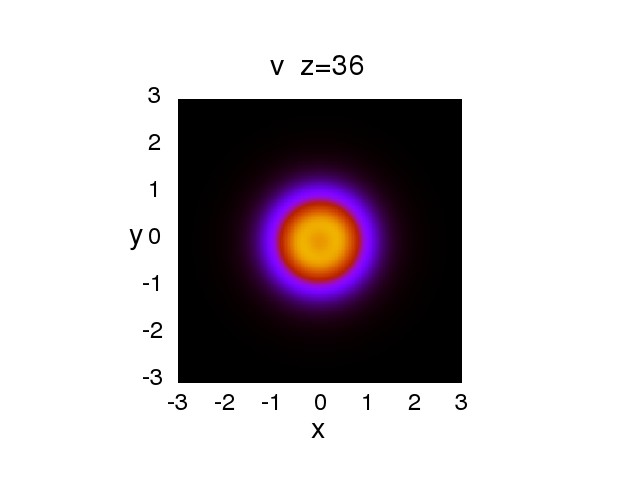} & %
\includegraphics[scale=0.17]{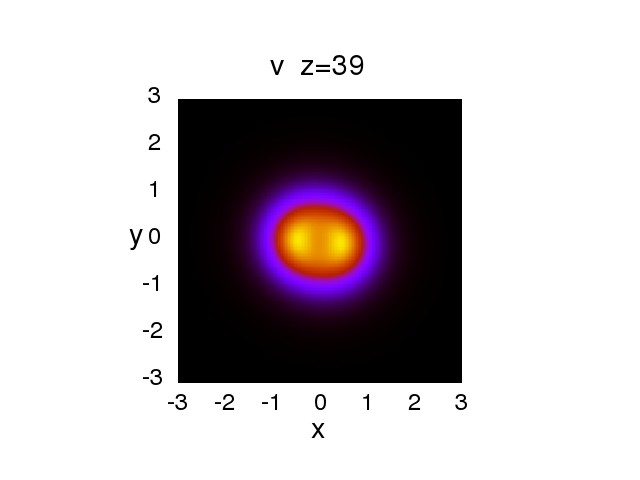} & %
\includegraphics[scale=0.17]{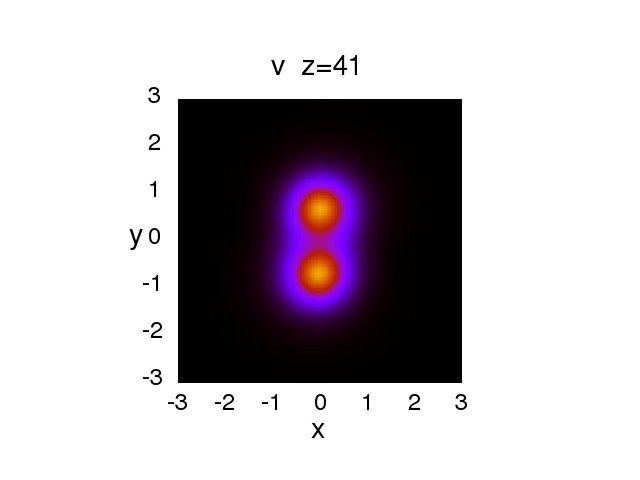} & %
\includegraphics[scale=0.17]{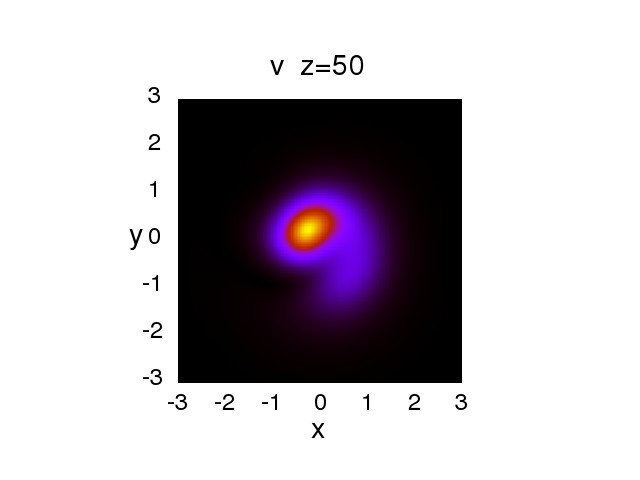} \\
\includegraphics[scale=0.17]{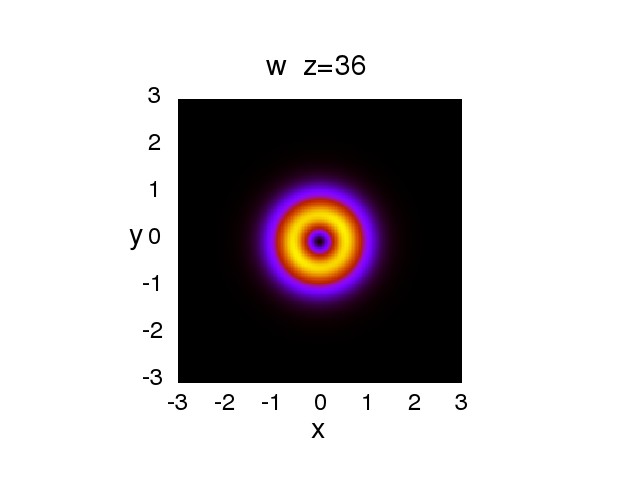} & %
\includegraphics[scale=0.17]{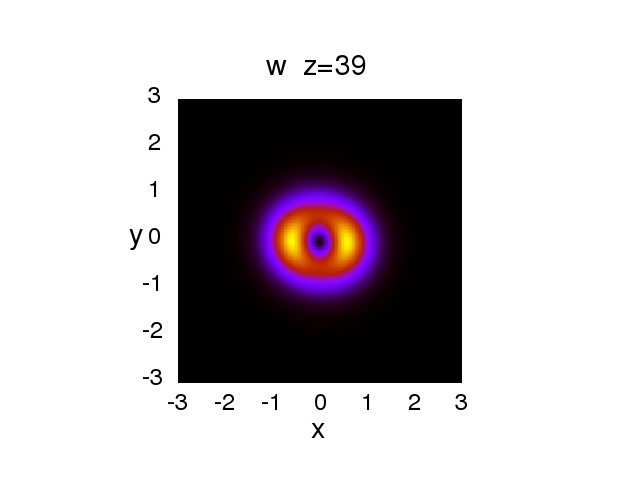} & %
\includegraphics[scale=0.17]{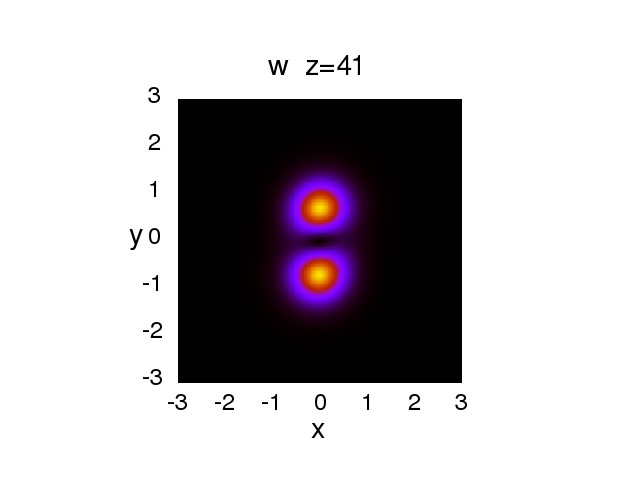} & %
\includegraphics[scale=0.17]{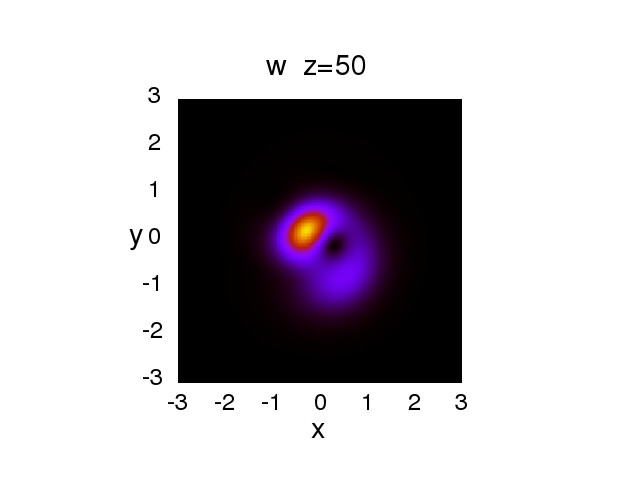} \\
&  &  &
\end{tabular}%
\caption{(Color online) The same as in Fig. \protect\ref{figm101_mu0.5}, but
for an unstable vortex ring corresponding to $k=1$ . }
\label{figm101_mu-1}
\end{figure}

\subsection{Full three-wave vortices}

The most general three-wave vortex state is built with $S_{u,v,w}=(
1,1,2) $. Families of these solutions were found for the symmetric
system, with $Q=0$ and $k_{u}=k_{v}\equiv k$ [see Eqs. (\ref{UV}) and (\ref%
{SH})] from a numerical solution of Eq. (\ref{1-1-2}). Their stability
eigenvalues were then computed using Eq. (\ref{eigen}), and the predicted
stability or instability was verified by direct simulations of Eqs. (\ref{u}%
)-(\ref{w}).

\begin{figure}[th]
\includegraphics[width=5.5cm,clip]{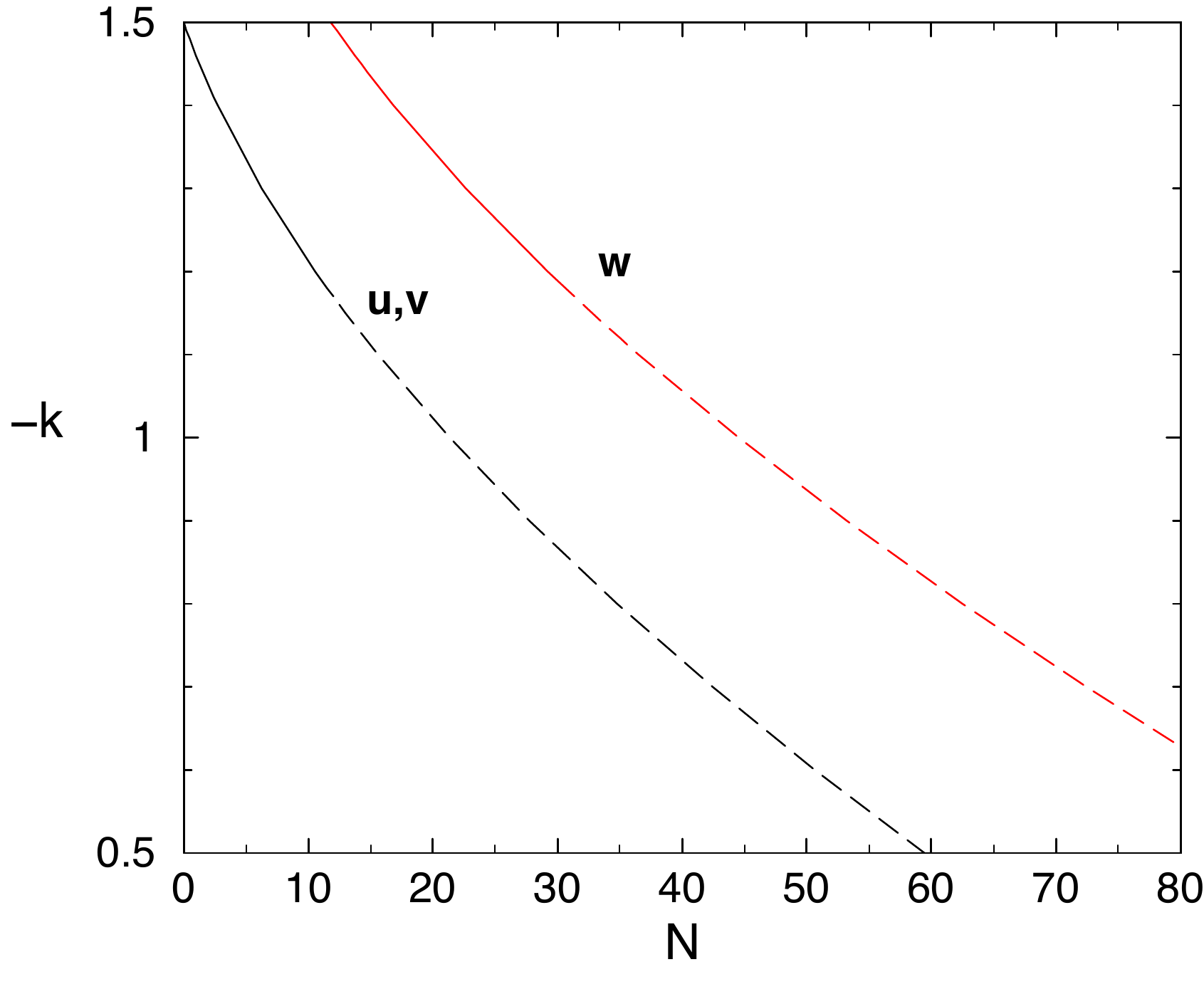} %
\includegraphics[width=5.5cm,clip]{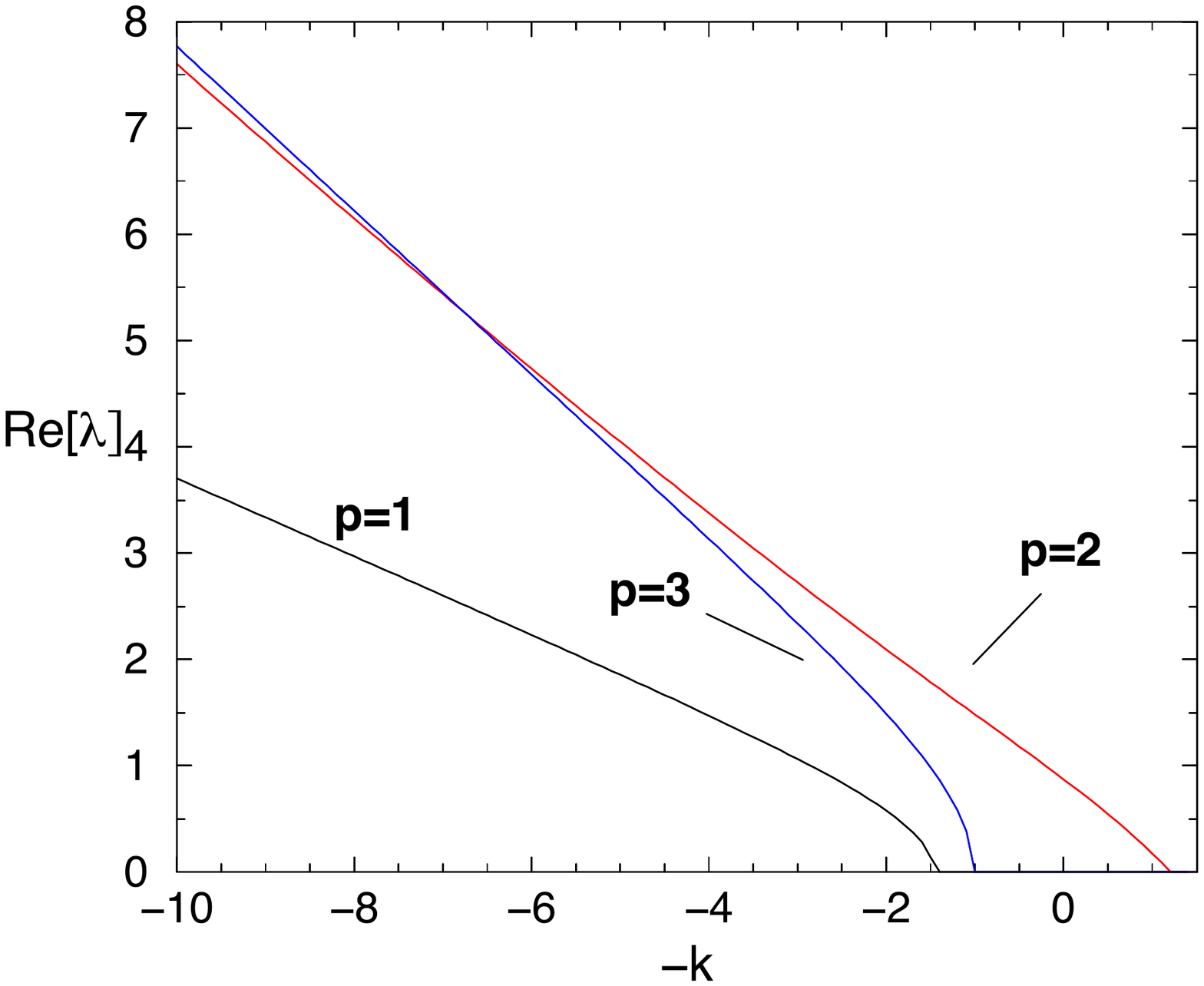}
\caption{(Color online) Left: The relation between the propagation constant
and integral powers of the three components in three-wave vortices with $%
S_{u,v,w}=(1,1,2) $, $\Omega =1$, $Q=0,q=0$. Solid and dashed
lines represent stable and unstable states, respectively. The transition
from the single-color state (the emergence of the FF components) occurs at $%
N_{w}=11.8$. The destabilization takes place at $k=-1.16$, $N_{u,v}=12.5$, $%
N_{w}=32.0$, the total power being $57.0$. Right: Instability eigenvalues
with the largest real part, as found from the numerical solution of
linearized equations (\protect\ref{eigen}). }
\label{figm112}
\end{figure}

A crucial difference of the three-wave vortices from the single-color
(SH-only), HV and SV states, which were considered above, is that the full
vortices have a well-defined stability area. A typical vortex family and its
stability are presented in Fig. \ref{figm112}. This figure explicitly
displays both the bifurcation, which generates the full vortex from the
corresponding single-color state, with $S_{w}=2$ (as shown above, the
bifurcation simultaneously destabilizes the single-color state), and the
point of the destabilization of the three-wave vortices.

\begin{figure}[th]
\includegraphics[width=5.5cm,clip]{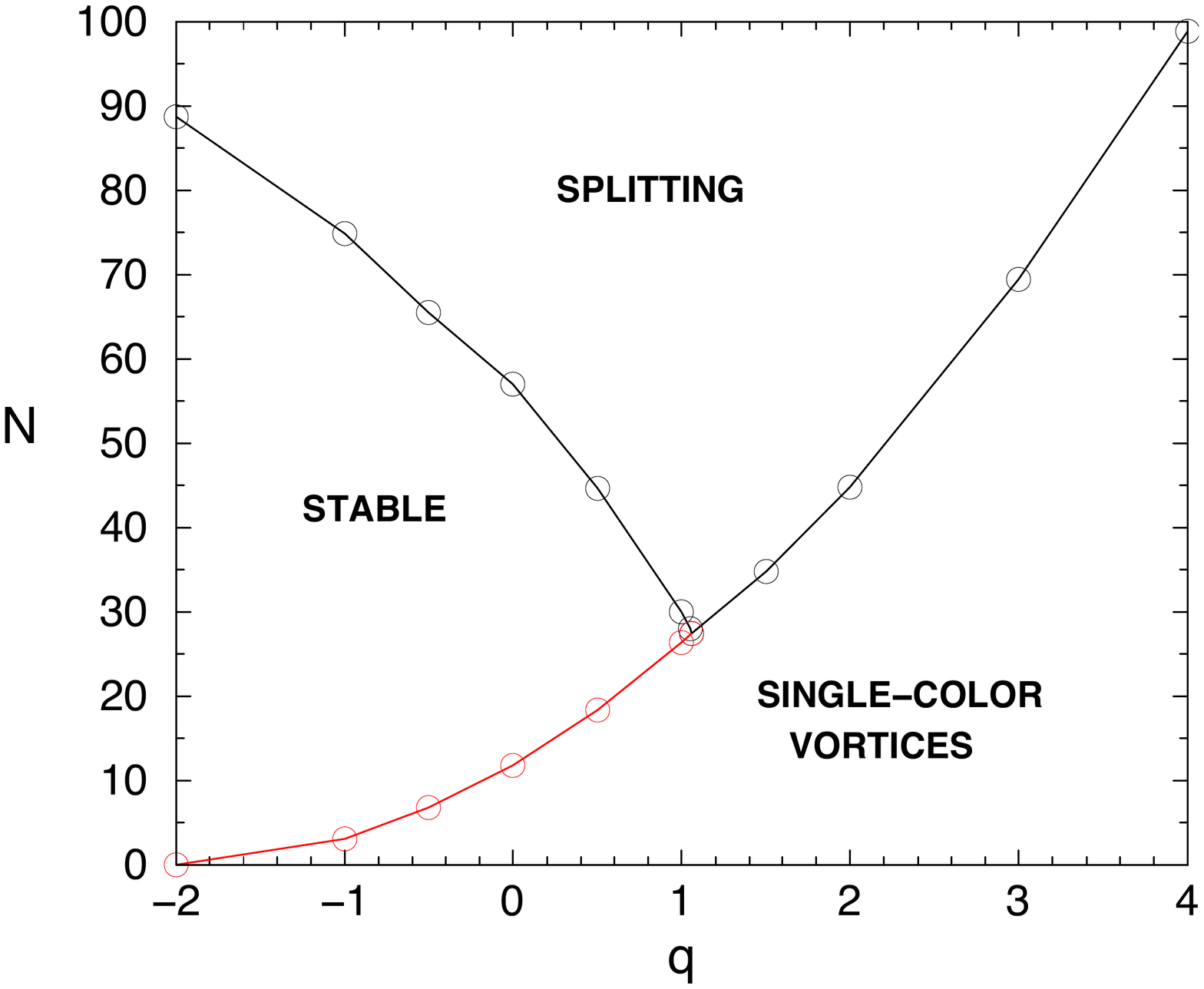} %
\includegraphics[width=5.5cm,clip]{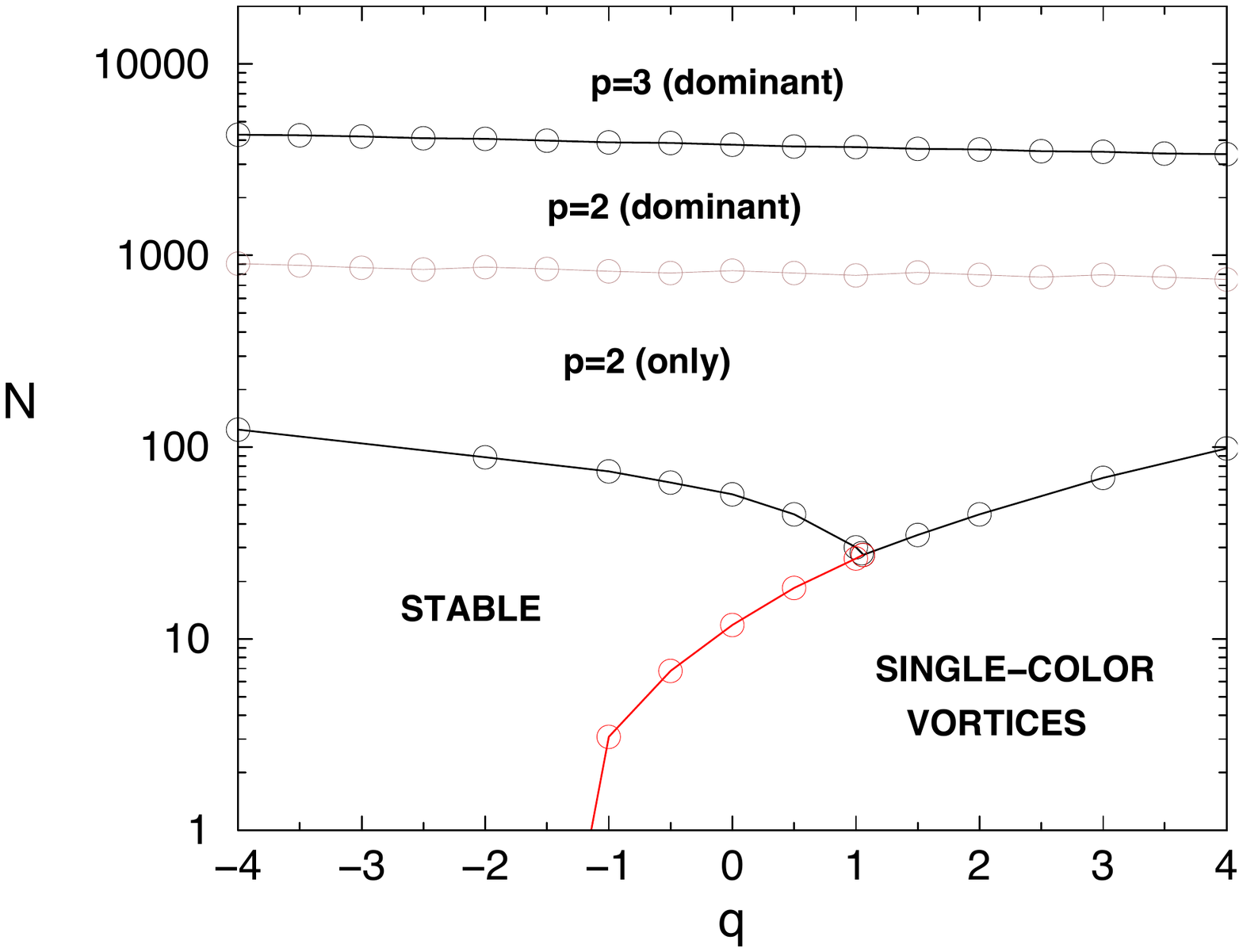}
\caption{(Color online) Left: The stability diagram for vortices with $%
S_{u,v,w}=(1,1,2) $, $\Omega =1$, $Q=0$, in the plane of the
mismatch parameter ($q$) and total power ($N$). Right: The same, on the
logarithmic scale of $N$. In this panel, dominant values of the perturbation
azimuthal index are displayed in the region of the splitting instability
(label ``only" implies that there is a single instability
eigenmode in the respective area).}
\label{figlog112}
\end{figure}

The results of the stability analysis are summarized, in the plane of the
phase-mismatch parameter, $q$, and total power, $N$, in Fig. \ref{figlog112}%
. The shape of the stability diagram is qualitatively similar to the one
which was recently reported, for the degenerate two-wave $\chi ^{(2)}$
system, in Ref. \cite{HS} (for the sake of the comparison, note that $q$ was
defined with the opposite sign in Ref. \cite{HS}). The bottom-right area in
the diagram is populated by the single-color (SH-only) vortices with $S_{w}=2
$, the bifurcation destabilizing the single-color vortex and replacing it by
the three-wave one occurring along the upper boundary of this area.

In the ``splitting" area labeled in Fig. \ref{figlog112},
the three-wave vortices are subject to an instability which splits them into
a set of fragments, the number of which is equal to the dominant (or single)
azimuthal index, $p$, of unstable perturbation modes, which is indicated in
the right panel of the figure. Further, in the region labeled
``$p=2$ (only)", the stable static three-wave vortices are
replaced by a robust dynamical regime, in the form of periodic splittings of
the vortex into two segments and their recombinations, as shown in Fig. \ref%
{fig9}. In the course of this periodic evolution, the vortical structure of
the mode is conserved. A similar dynamical regime, in the form of periodic
splittings and recombinations, is known in the 2D GP equation with the cubic
self-attractive nonlinearity and HO trapping potential \cite{cubic-in-trap5}%
. On the other hand, the instability-induced splitting the vortex into a set
of three fragments is irreversible, as shown in Fig. \ref{figm112high}.

\begin{figure}[th]
\begin{tabular}{cccc}
\includegraphics[scale=0.17]{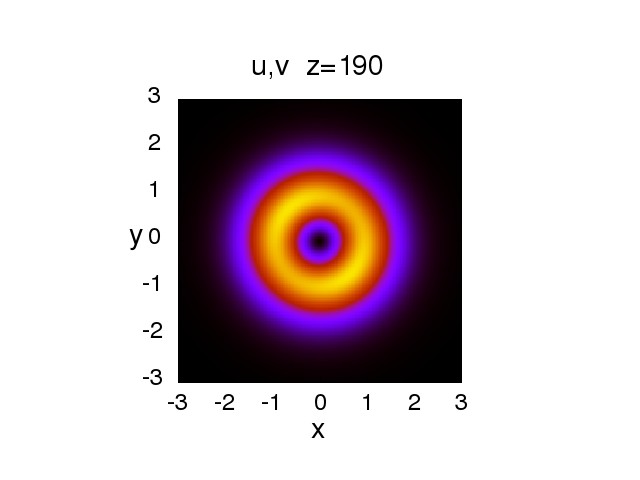} & %
\includegraphics[scale=0.17]{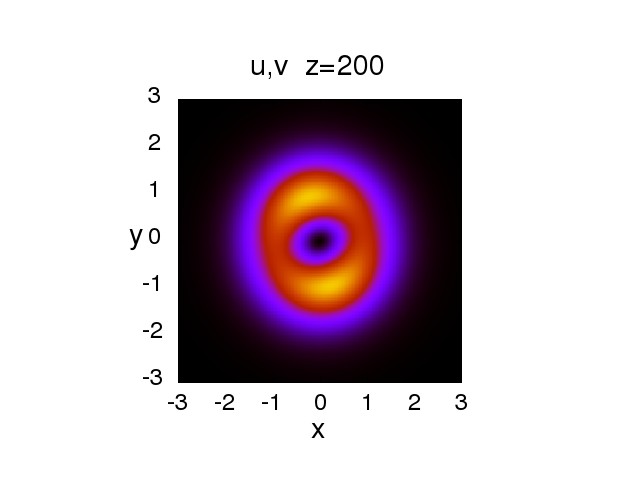} & %
\includegraphics[scale=0.17]{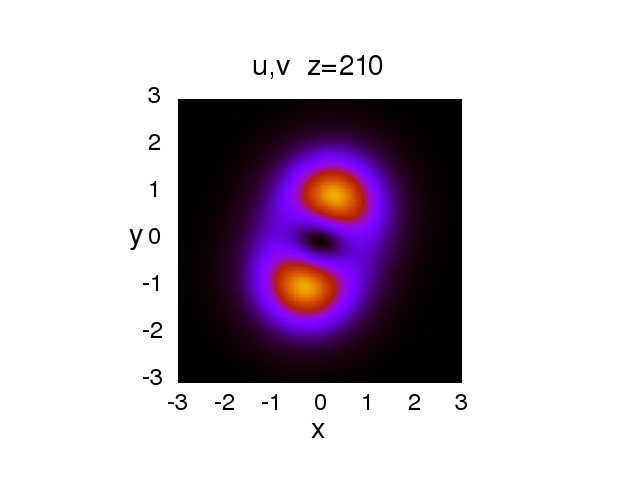} & %
\includegraphics[scale=0.17]{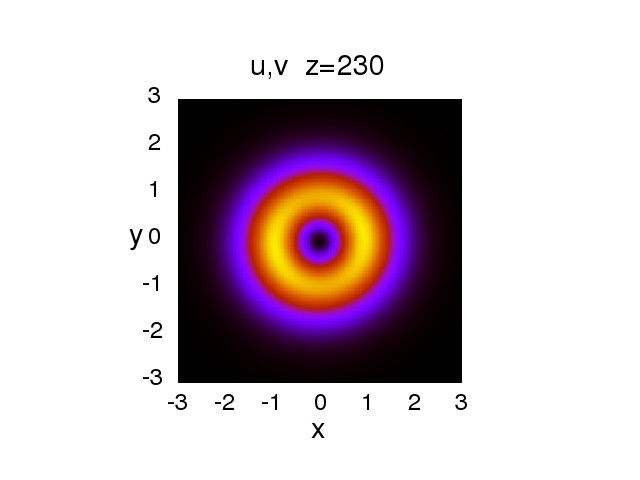} \\
\includegraphics[scale=0.17]{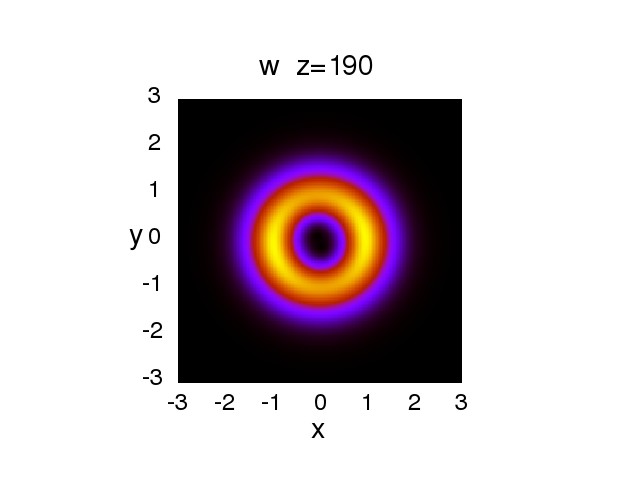} & %
\includegraphics[scale=0.17]{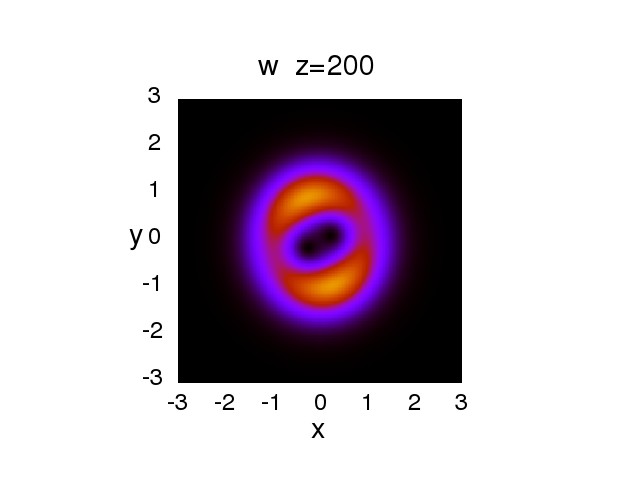} & %
\includegraphics[scale=0.17]{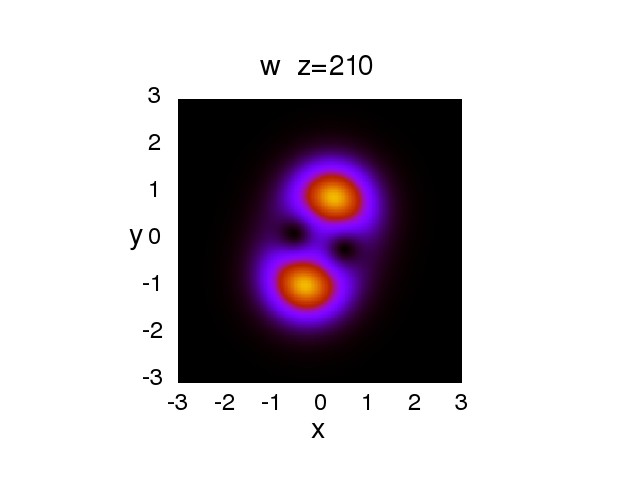} & %
\includegraphics[scale=0.17]{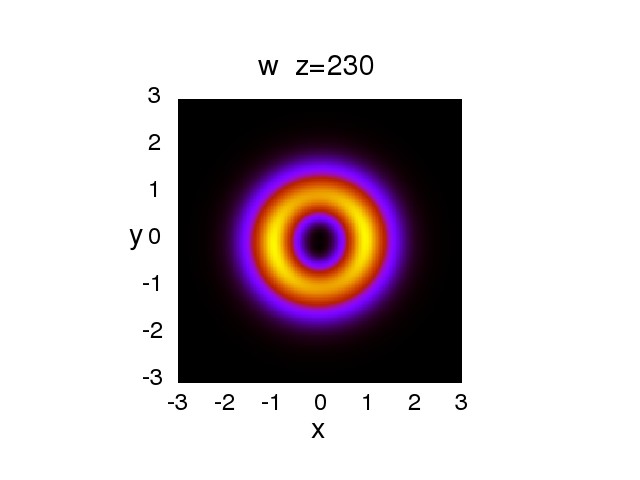} \\
&  &  &
\end{tabular}%
\caption{(Color online) The evolution of the densities of the $u,v,w$ fields
in the regime of periodic splittings and recombinations of the three-wave
vortex, with $S_{u,v,w}=(1,1,2) $, $\Omega =1$, $Q=0,q=0$, $%
k=-1$, $N_{u,v}=21.4$, $N_{w}=44.4$,\ the total power being $N=87.2$. }
\label{fig9}
\end{figure}

\begin{figure}[th]
\begin{tabular}{cccc}
\includegraphics[scale=0.17]{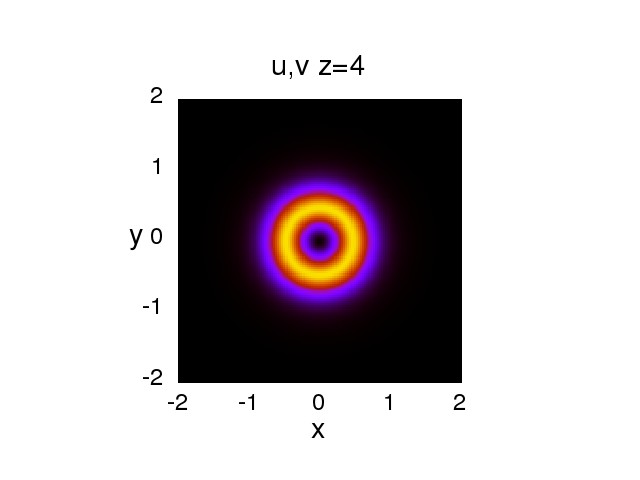} & %
\includegraphics[scale=0.17]{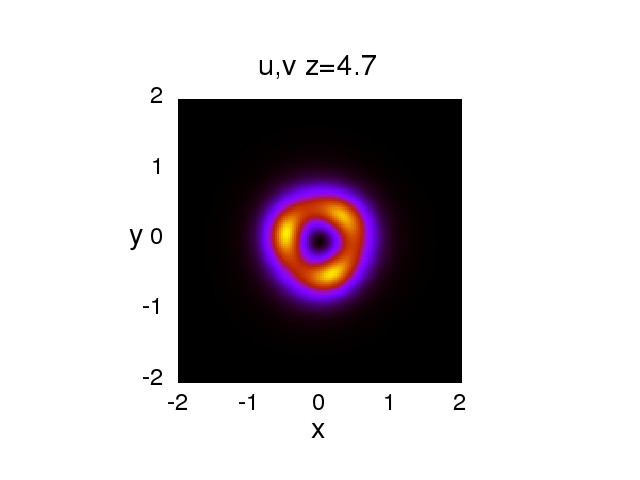} & %
\includegraphics[scale=0.17]{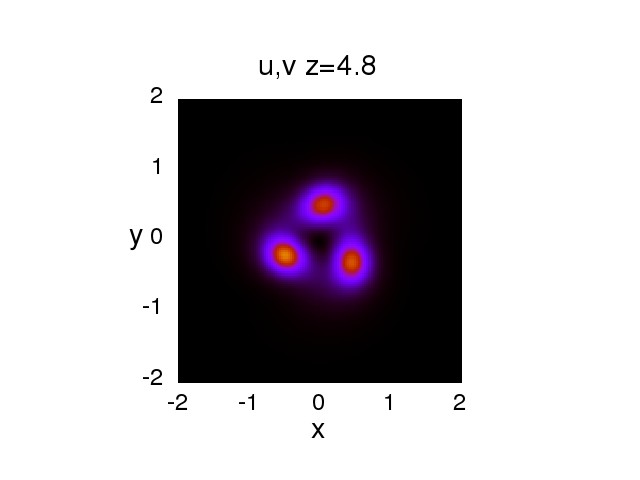} & %
\includegraphics[scale=0.17]{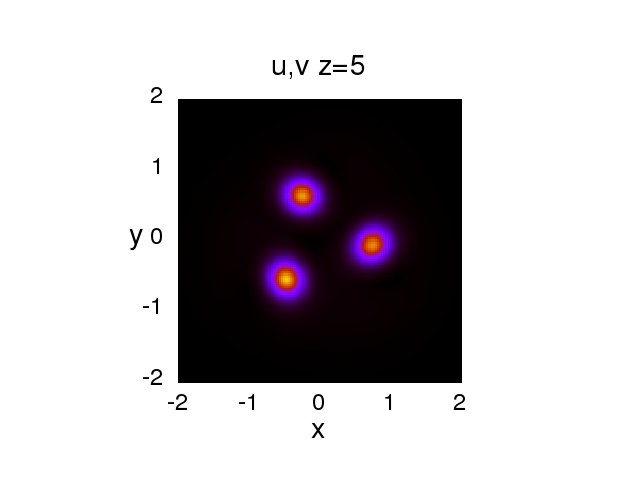} \\
\includegraphics[scale=0.17]{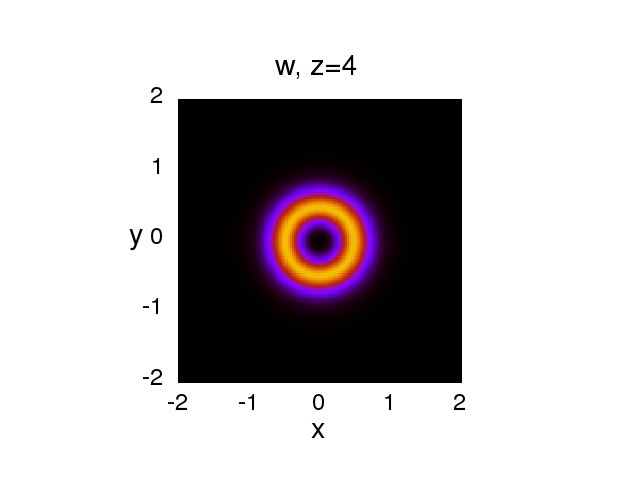} & %
\includegraphics[scale=0.17]{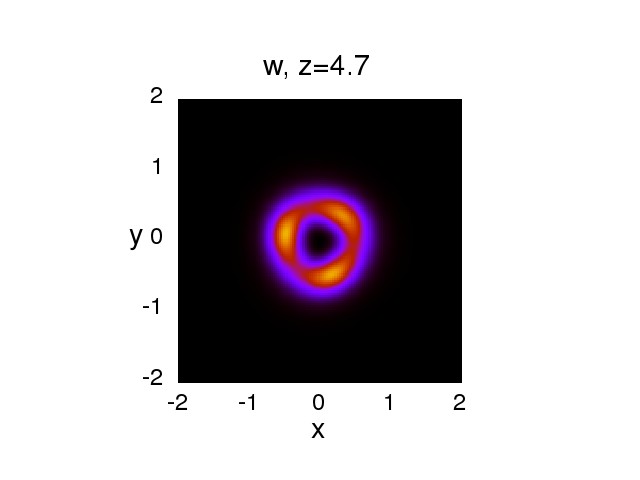} & %
\includegraphics[scale=0.17]{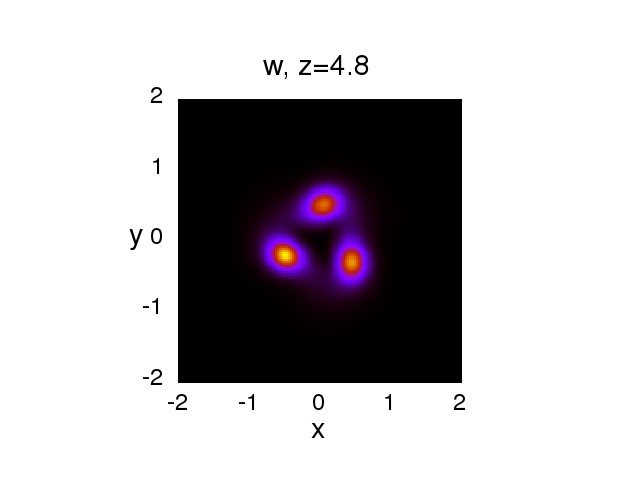} & %
\includegraphics[scale=0.17]{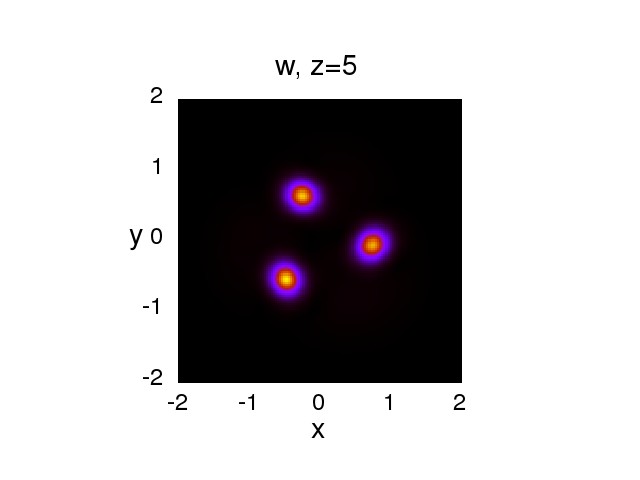} \\
&  &  &
\end{tabular}%
\caption{(Color online) An example of the irreversible splitting of the
three-wave vortex with $S_{u,v,w}=(1,1,2) $. Top: The
evolution of densities of the FF fields, $u$ and $v$, for $\Omega =1$, $%
Q=0,q=0$, $k=10$, $N_{u,v}=1800$, $N_{w}=2142$, the total power being $%
N=5741$. Bottom: The same but for the SH$\ $field, $w$. }
\label{figm112high}
\end{figure}

\begin{figure}[th]
\includegraphics[width=5.5cm,clip]{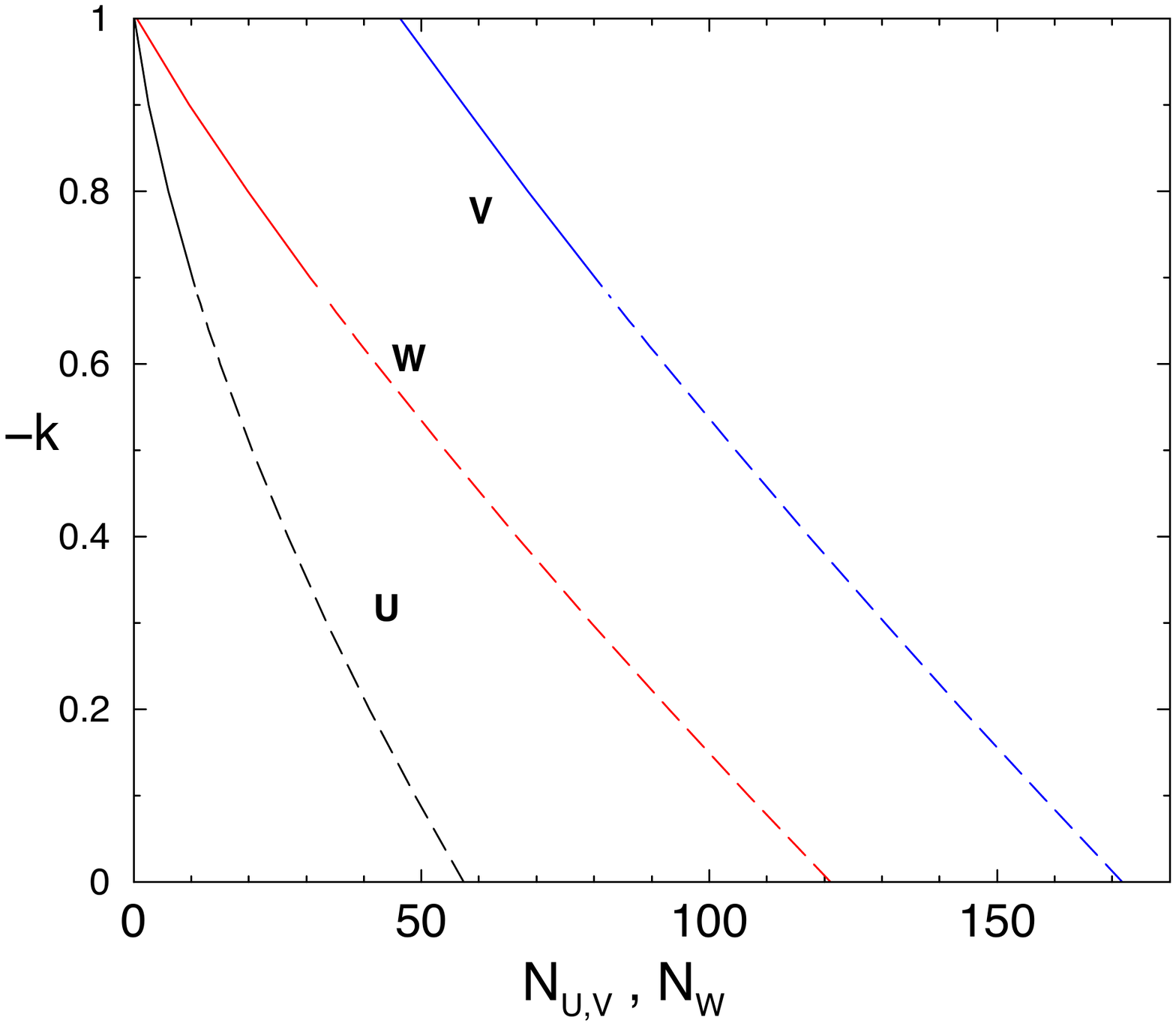} %
\includegraphics[width=5.5cm,clip]{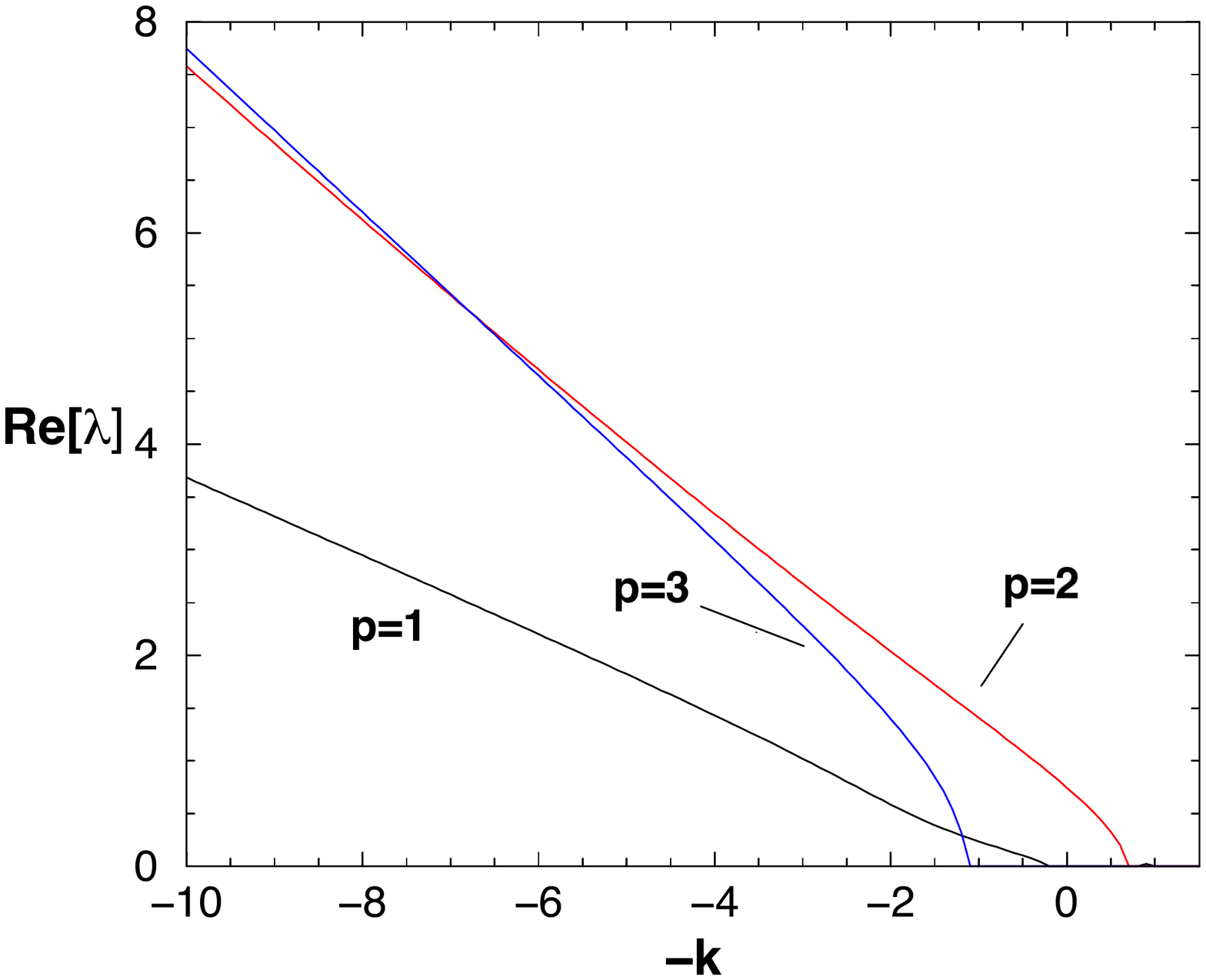}
\caption{(Color online) The same as in the right panel of Fig. \protect\ref%
{figm112}, but with $Q=1$ in Eqs. (\protect\ref{u})-(\protect\ref{v}).}
\label{figm112Q1q0}
\end{figure}

Finally, the effect of the birefringence of the stability of the three-wave
vortices, represented by $Q=1$ in Eqs. (\ref{u})-(\ref{v}), was briefly
considered too. As shown in Fig. \ref{figm112Q1q0}, the birefringence terms
make the stability area somewhat larger. This result can be explained by the
fact that the birefringence renders the system less coherent, while the
splitting instability of the vortices is a result of highly coherent $\chi
^{(2)}$ interactions.

\section{Conclusion}

This work aimed to explore the possibility of the stabilization of various
2D three-wave modes supported by the $\chi ^{(2)}$ interactions in the
combination with the isotropic HO (harmonic-oscillator) trapping potential.
The existence and stability of the modes is determined by powers and
vorticities of the three components and the mismatch of the $\chi ^{(2)}$
system ($q$). First, using both numerical computations and the VA\
(variational approximation), stability boundaries were identified for the
fundamental (zero-vorticity) and vortical single-color states, in which only
the SH (second-harmonic) component is present. On the contrary to the usual
assumption that the single-color SH modes are subject to the parametric
instability against perturbations in the FF (fundamental-frequency) fields,
we have found\ that they are \emph{stable} below the respective critical
values of the total power. Next, HV (hidden-vorticity) and SV\ (semi-vortex)
states, with vorticities, respectively, $\pm 1$ or $1$ and $0$ in the two FF
components, were found to be always unstable. The stability region has been
identified for the full three-wave vortices. Furthermore, adjacent to it is
the region which features the robust dynamical regime of periodic splitting
into two fragments and their recombination into the original vortex.

The analysis reported in the present work can be extended in other
directions. One possibility is to analyze asymmetric three-wave modes, with
unequal propagation constants in the two components of the FF field. It may
also be interesting to construct self-trapped three-wave $\chi ^{(2)}$ modes
supported by a periodic (lattice) potential, instead of the HO, cf. Ref.
\cite{chi2-in-lattice}. Lastly, a challenging possibility is to construct
three-dimensional three-wave ``light bullets" supported by
the HO trapping potential, cf. Ref. \cite{HS-3D} where this was done for the
degenerate two-wave system.

A.G. would like to thank the Brazilian funding
agencies Funda\c{c}\~ao de Amparo \`a Pesquisa do Estado de S\~ao Paulo
and Conselho Nacional de Desenvolvimento Cient\'ifico e
Tecnol\'ogico. B.A.M. acknowledges a visitor's grant
provided by South American Institute for Fundamental Research (Sao
Paulo).


\begin{thebibliography}{99}
\bibitem{review0} Stegeman G I, Hagan D J  and Torner L 1996
{\it Opt. Quant. Electr.} \textbf{28} 1691-1740

\bibitem{review1} Etrich C, Lederer F, Malomed B A, Peschel T and
Peschel U 2000
{\it Progr. Opt.} \textbf{41} 483-568

\bibitem{review2} Buryak A V, Di Trapani P, Skryabin D V and S.
Trillo S. 2002
{\it Phys. Rep.} \textbf{370} 63-235

\bibitem{review} Malomed B A, Mihalache D, Wise F and Torner L 2005
{\it J. Optics B: Quant. Semicl. Opt.} \textbf{7}, R53-R72

\bibitem{KA} Kivshar Y S and Agrawal G P 2003, \textit{Optical Solitons: From
Fibers to Photonic Crystals} (San Diego, CA: Academic)

\bibitem{Du} Du F, Lu Y W and Wu S T 2004
{\it Appl. Phys. Lett.} \textbf{85} 2181-2183

\bibitem{Luan} Luan F, George A K, Hedeley T D, Pearce G J,
Bird D M, Knight J C and Russell P S J 2004
{\it Opt. Lett.} \textbf{29}  2369-2371

\bibitem{Rubenchik} Kanashov A A and Rubenchik A M 1981
{\it Physica D} \textbf{4}, 122-134

\bibitem{first-exper} Torruellas W E, Wang Z, Hagan D J,
VanStryland E W, Stegeman G I, Torner L and Menyuk C R 1995
{\it Phys. Rev. Lett.} \textbf{74} 5036-5039

\bibitem{Drummond} Malomed B A, Drummond P, He H, Berntson A,
Anderson D and Lisak M 1997
{\it Phys. Rev. E} \textbf{56} 4725-4735

\bibitem{Wise1} Liu X, Qian L J and Wise F W 1999
{\it Phys. Rev. Lett.} \textbf{82} 4631-4634

\bibitem{Wise2} Liu X, Beckwitt K and Wise F 2000
{\it Phys. Rev. E} \textbf{62} 1328-1340

\bibitem{half} Bovino F A, Braccini M and Sibilia C 2011
{\it J. Opt. Soc. Am. B} \textbf{28} 2806-2811

\bibitem{splitting1} Firth W J and Skryabin D V 1997
{\it Phys. Rev. Lett.} \textbf{79} 2450-2453

\bibitem{splitting2} Torner L and Petrov D V 1997
{\it Electron. Lett.} \textbf{33} 608-610

\bibitem{splitting4} Skryabin D V and Firth W J 1998
{\it Phys. Rev. E} \textbf{58} R1252-R1255

\bibitem{splitting5} Torres J P, Soto-Crespo J M, Torner L and Petrov D V 1998
{\it J. Opt. Soc. Am. B} \textbf{15} 625-627

\bibitem{splitting-exp} Petrov D V, Torner L, Martorell J, Vilaseca R,
Torres J P and Cojocaru C 1998
{\it Opt. Lett.} \textbf{23} 1444-1446

\bibitem{splitting-3W} Torres J P, Soto-Crespo J M, Torner L and Petrov D V 1998
{\it Opt. Commun.} \textbf{149} 77-83

\bibitem{splitting-3W2} Molina-Terriza G, Wright E M and Torner L 2001
{\it Opt. Lett.} \textbf{26} 163-165

\bibitem{Kruglov} Kruglov V I, Logvin Y A and Volkov V M 1992
{\it J. Mod. Opt.} \textbf{39} 2277-2291

\bibitem{BEC} Pethick C J and Smith H 2008 \textit{Bose-Einstein condensate
in dilute gas }(Cambridge University Press: Cambridge)

\bibitem{cubic-in-trap1} Dalfovo F and  Stringari S 1996
{\it Phys. Rev. A} \textbf{53} 2477-2485

\bibitem{cubic-in-trap2} Dodd R J 1996
{\it J. Res. Natl. Inst. Stand. Technol.} \textbf{101} 545-552

\bibitem{cubic-in-trap3} Alexander T J and Berg\'{e} L 2002
{\it Phys. Rev. E} \textbf{65} 026611

\bibitem{cubic-in-trap4} Carr L D and Clark C W 2006
{\it Phys. Rev. Lett.} \textbf{97} 010403

\bibitem{cubic-in-trap5} Mihalache D, Mazilu D, Malomed B A and Lederer F 2006
{\it Phys. Rev. A} \textbf{73} 043615

\bibitem{cubic-in-trap6} Carr L D and Clark C W 2006
{\it Phys. Rev. A} \textbf{74} 043613

\bibitem{cubic-in-trap7} Herring G, Carr L D, Carretero-Gonz\'{a}lez R,
Kevrekidis P G and Frantzeskakis D J 2008
{\it Phys. Rev. A} \textbf{77} 043607

\bibitem{semi} Sakaguchi H, Li B and Malomed B A 2014
{\it Phys. Rev. E} \textbf{89} 032920

\bibitem{HS} Sakaguchi H and Malomed B A 2012,
{\it J. Opt. Soc. Am. B} \textbf{29} 2741

\bibitem{HS-3D} Sakaguchi H and B. A. Malomed 2013
{\it Opt. Exp.} \textbf{21} 9813-9823

\bibitem{BEC0} Drummond P D, Kheruntsyan K V and He H 1998
{\it Phys. Rev. Lett.} \textbf{81} 3055-3058

\bibitem{BEC1} Heinzen D J, Wynar R, Drummond P D and
Kheruntsyan K V 2000
{\it Phys. Rev. Lett.} \textbf{84} 5029-5033

\bibitem{BEC2} Hope J J and Olsen M K 2001
{\it Phys. Rev. Lett.} \textbf{86} 3220-3223

\bibitem{BEC4} Hornung T, Gordienko S, de Vivie-Riedle R and Verhaar 2002
{\it Phys. Rev. A} \textbf{66} 043607

\bibitem{Lluis} Torres J P, Soto-Crespo J M, Torner L and
Petrov D V 1998
{\it Opt. Commun.} \textbf{149} 77-83

\bibitem{hidden0} Desyatnikov A S, Mihalache D, Mazilu D, Malomed B A,
Denz C and Lederer F 2005
{\it Phys. Rev. E} \textbf{71} 026615

\bibitem{Herve} Leblond H, Malomed B A and Mihalache D 2005
{\it Phys. Rev. E} \textbf{71} 036608

\bibitem{Anton} Desyatnikov A S, Mihalache D, Mazilu D, Malomed B A
and Lederer F 2007
{\it Phys. Lett. A} \textbf{364} 231-234

\bibitem{hidden} Brtka M, Gammal A and Malomed B A 2010
{\it Phys. Rev. A} \textbf{82} 053610

\bibitem{Panos} Kevrekidis P G and Pelinovsky D E 2006
{\it Proc. R. Soc. A} \textbf{462} 2671-2694

\bibitem{Leykam} Leykam D, Malomed B and Desyatnikov A S 2013
{\it J. Optics} \textbf{15} 044016

\bibitem{Fangwei}
Ye F, Wang J, Dong L, and Li Y 2004,
{\it Opt. Commun.} \textbf{230} 219-223

\bibitem{chi2-in-lattice} Xu Z Y, Kartashov Y V, Crasovan L C, Mihalache D and Torner L 2005
{\it Phys. Rev. E} \textbf{71} 016616
\end{thebibliography}
\end{document}